\documentclass[sn-mathphys,Numbered]{sn-jnl}

\usepackage{graphicx}%
\usepackage{multirow}%
\usepackage{amsmath,amssymb,amsfonts}%
\usepackage{amsthm}%
\usepackage{mathrsfs}%
\usepackage[title]{appendix}%
\usepackage{xcolor}%
\usepackage{textcomp}%
\usepackage{manyfoot}%
\usepackage{booktabs}%
\usepackage{algorithm}%
\usepackage{algorithmicx}%
\usepackage{algpseudocode}%
\usepackage{listings}%
\usepackage{makecell}
\usepackage{bm}
\usepackage{soul}
\usepackage[T1,T2A]{fontenc}
\usepackage{appendix}
\usepackage{ulem}
\usepackage{subfig}

\newcommand{\be}{\begin{equation}}
\newcommand{\ee}{\end{equation}}
\newcommand{\bs}{\begin{split}} 
\newcommand{\bea}{\begin{eqnarray}}
\newcommand{\eea}{\end{eqnarray}}
\newcommand\ringring[1]{%
  {
   \mathop{\kern0pt #1}\limits^{
     \vbox to-1.85ex{
       \kern-2ex 
       \hbox to 0pt{\hss\normalfont\kern.1em \r{}\kern-.45em \r{}\hss}%
       \vss 
     }
   }
  }
}

\begin{document}

\title[Article Title]{The exponential metric: traversable wormhole and possible identification of scalar background}

\author[1]{\fnm{Eduard} \sur{Mychelkin}}\email{mychelkin@fai.kz}

\author[1,2]{\fnm{Gulnara} \sur{Suliyeva}}\email{suliyeva@fai.kz}

\author*[1]{\fnm{Maxim} \sur{Makukov}}\email{makukov@fai.kz}

\affil*[1]{\orgname{Fesenkov Astrophysical Institute}, \orgaddress{\street{Observatory 23}, \city{Almaty}, \postcode{050020}, \country{Kazakhstan}}}

\affil[2]{\orgname{Al-Farabi Kazakh National University}, \orgaddress{\street{Al-Farabi av. 71}, \city{Almaty}, \postcode{050040}, \country{Kazakhstan}}}


\abstract{

The static antiscalar solution of the Einstein-Klein-Gordon equations in the form of the Papapetrou exponential metric had been interpreted as a traversable wormhole with a throat at \textit{r=M}. We aim to search for the effects which could be associated with this scale and only find that the topological Gauss-Bonnet invariant swaps sign, and the value of the Keplerian frequency for circular geodesics becomes singular. At the same time, the geometric invariants of the curvature tensor have extremal values at the scale twice less than that of the throat, revealing new physical effects. In particular, the Ricci scalar at \textit{r=M/2} (rather than \textit{r=M}) is associated with the extremal values of thermodynamic characteristics of the scalar background. This approach in combination with the antiscalar static limit of the Einstein-Maxwell equations suggests the interpretation of the scalar background as a stable medium with a stiff equation of state, formed by the neutral superposition of ambient quasistatic electric fields.

}

\keywords{antiscalar solution, traversable wormholes, curvature invariants, scalar thermodynamics, quasistatic electric fields}

\maketitle

\section{Introduction}\label{sec1}

In general relativity (GR), in the class of spherically symmetric static spacetimes, an exclusive role belongs to the vacuum Schwarzschild metric typically represented in curvature coordinates.  An observationally viable alternative is the exponential Papapetrou metric \cite{papa54}. It represents the solution of the Einstein-scalar and Klein-Gordon equations in the stable anti-scalar regime; it is obtained directly in isotropic coordinates and does not contain horizons \cite{2018PhRvD..98f4050M}. 

The antiscalar regime implies a sign opposite to the canonical one for the scalar energy-momentum tensor (EMT, here massless) in the Einstein equations. Due to the non-canonical form of the Einstein-Gilbert part of the Lagrangian, in general, it cannot \textit{a priori} prescribe this sign, and thereby it should be determined primarily from the compliance with observations. We prefer to avoid using such terms as ``ghost'' or ``phantom'' for the source field, since there is a single standard scalar EMT (obeying exotic equations of state), and the term ``phantom'' has already been adopted for the cosmological scalar field with the state parameter $w<-1$.

As for the complience with observations, two independent studies \cite{2018PhRvD..98f4050M} and \cite{2018PhRvD..Visser} have addressed this issue in terms of effects such as photon spheres and innermost stable circular orbits, and the results agree. In \cite{2018PhRvD..98f4050M}, we also examined photon shadow characteristics and geodetic precession, finding no contradiction with current observational data. In the weak-field approximation, the predictions of the vacuum and antiscalar models are practically indistinguishable. In \cite{2024GReGr..56...44M}, we analyzed the perihelion shift per century for highly eccentric S-cluster star orbits in the antiscalar regime, showing that the distinction from the vacuum prediction is negligible within the accuracy of current instruments. However, a more pronounced effect was predicted for the orbit of the S62 star. Additionally, an evaluation in \cite{2024GReGr..56...44M} of the strong-field shadow effect \cite{2019ApJ...875L...1E} yielded a notable increase in shadow size (over 5\%) compared to the vacuum solution. This excess might serve as a potentially observable signature, provided the central mass estimate is refined.

The paper \cite{2018PhRvD..Visser} gives the interpretation of the Papapetrou exponential metric as a traversable wormhole (TWH) (the Schwarzschild metric is a non-traversable wormhole due to presence of the horizon). The essential feature of the exponential metric as a TWH model is a throat at the radial coordinate \textit{r=M} which connects two distinct regions of spacetime.

In this paper, we aim to search for the geometric and physical phenomena related to the given TWH scale (this concerns the topological Gauss-Bonnet invariant and geodesic equations for the test particles) as well as to other critical scales related to different geometric invariants of the Riemann-Christoffel curvature tensor leading to corresponding thermodynamic features of the scalar field. 

Another aim for us here is to compare the effects in scalar field with those in the Schwarzschild vacuum, important for the existence problem of the scalar background from observational viewpoint. Then, taking into account the thermodynamic consideration in connection with the antiscalar state of the Einstein-Maxwell equations, we outline a possible direction to approach the problem of the origin of the fundamental scalar background.

\section{Set-up}\label{setup}

We begin by writing down the Schwarzschild and Papapetrou metrics in isotropic coordinates \cite{2024GReGr..56...44M}:
\begin{equation}
ds^2 (r)= B(r) dt^2 - D(r) \left( dr^2 +  r^2 d\theta^2
       +  r^2 \sin^2 \theta d\phi^2\right),
\label{isotrop}
\end{equation}
where, in case of the Papapetrou metric,
\begin{equation} \label{Pp}
B=D^{-1}=e^{-2M/r}=e^{-2\varphi(r)},
\end{equation} 
while for the Schwarzschild case
\begin{equation} \label{Ss}
B=\left(\frac{1-\frac{M}{2r}}{1+\frac{M}{2r}}\right)^2=\left(\frac{1-\frac{\varphi(r)}{2}}{1+\frac{\varphi(r)}{2}}\right)^2,\quad D=\left(1+\frac{M}{2r}\right)^4=\left(1+\frac{\varphi(r)}{2}\right)^4.
\end{equation}
The explicit exponential metric for the Einstein-Klein-Gordon system,
\begin{equation} 
ds^2(r) =  e^{-2\varphi(r)} dt^2 - 
 e^{2\varphi(r)} (dr^2 + r^2 d\theta^2 + r^2 \sin^2\theta
d\phi^2), \quad \varphi(r)=M/r,
\label{Pap}
\end{equation}
represents the Papapetrou solution of the Einstein-scalar equations in the
antiscalar regime, and the scalar field $\varphi(r)$ satisfies the massless Klein-Gordon equation:
\begin{equation} 
{G}_{\mu\nu} = \epsilon \varkappa {T}_{\mu\nu}^{\text{SF}}\left( \varphi \right), \quad {T}^{\text{SF}}_{\mu\nu}(\varphi) = \frac{1}{4 \pi} \left( \varphi_\mu
\varphi_\nu - \frac{1}{2} {g}_{\mu\nu} \varphi^\alpha \varphi_\alpha
\right), \,\,\,\, \varphi_\mu \equiv \partial_\mu \varphi,
\label{EinEq}
\end{equation}
and
\begin{equation}
\square \varphi \equiv  {\varphi_{;\alpha}}^{;\alpha} = 0, 
\label{KG}
\end{equation}
where $\varkappa=8 \pi$ (in units $G=c=1$), and $\epsilon=(+1, 0, -1)$
is the indicator of the relation to geometry for scalar, vacuum and
antiscalar modes, respectively.

The basic equations \eqref{EinEq} and \eqref{KG} must be supplemented with a system of geodesic equations. For spacetimes with spacelike ($\xi_{(\phi)} = \partial/\partial \phi$) and timelike ($\xi_{(t)} = \partial/\partial t$) killing vectors, the constants of motion arise as projections onto the 4-velocity of the test particles $u^{\alpha}$. These are angular momentum (at $\theta = \pi/2$),
\begin{equation}\label{defL}
\tilde L \equiv u_a \xi^a_{(\phi)} = u_\phi 
= g_{\phi\phi} \frac{d\phi}{d\tau}
= - D r^2 \dot{\phi}
= \mbox{const}, 
\end{equation}
and energy,
\begin{equation} \label{defE}
\tilde E \equiv u_a \xi^a_{(t)} = u_t 
= g_{tt} \frac{dt}{d\tau}
=  B \dot{t}
= \mbox{const},
\end{equation}
both per unit mass, as denoted here with tildes. From this, in
particular, one immediately obtains the expression for the angular
velocity (up to a sign),
\begin{equation} \label{Om}
\Omega (r)\, \dot{=} \, \frac{\dot{\phi}}{\dot{t}} =\frac{B\tilde{L}}{Dr^2\tilde{E}} \,\, ,
\end{equation} 
to which we will resort later.

\section{Double role of TWH critical scales}\label{sec:conception}

If, following \cite{2018PhRvD..Visser}, within the Papapetrou spacetime \eqref{Pap}, we identify the minimum of area-function $A(r)$ for equipotential surfaces with $g_{00}=const$, $t=const$:
\begin{equation}\label{area}
    A(r) = \int_0^{2\pi} \int_0^{\pi} \sqrt{g_{\theta\theta}g_{\phi\phi}}d\theta d\phi =4\pi r^2 e^{2M/r}, \quad A'(r)=0 \quad \Rightarrow \quad r=M,
\end{equation}
then, obviously, from the condition $A'(r)=0$ the minimal value of the radial
coordinate $r=M$ will be found, at which the ``flare-out'' requirement
$A''(r=M)>0$ is also satisfied. This results in the TWH throat, in the sense of Morris and Thorne \cite{1988AmJPh..56..395M}.

It is convenient to introduce the effective (``physical'') radius $\rho(r)$ such that the ``curved'' area \eqref{area} assumes the standard (in terms of cylindrical Euclidean space) form: $A(r) \to A(\rho)=4\pi \rho^2(r)$. So, in the Papapetrou metric at the critical scale
$r=M$ we get a minimum area $A_0=4\pi e^2M^2$ on the set of spheres of effective radius $\rho(r)\dot{=}{re^{M/r}}$:
\begin{equation} \label{minAr}
A_0=\text{min}\{A(\rho)\}=\text{min} \{4\pi \rho^2 (r)\}=4\pi \rho_0^2 =4\pi(eM)^2.
\end{equation}
In this terms, the new minimal value is:
\begin{equation} \label{minR}
\rho_0=\rho(r)|_{r=M}\equiv \rho(M) = eM = 2.71828M 
\end{equation}
which means the ``physical'' scale of the throat to be distinguished from the initial ``coordinate'' scale,\footnote{Such terminology corresponds to \cite{2024Univ...10..328G}, for the TWH throat within the de Sitter cosmology. Note also that the method of \cite{2024Univ...10..328G} is based on a formal transition to a purely imaginary scalar field, exactly as in \cite{2018PhRvD..98f4050M}.} $r=M$. 

Note that the same relation \eqref{minR} might be obtained in the framework of curvilinear integral for a circle in spacetime in static coordinates ($t,r,\theta,\phi$) indexed as ($0,1,2,3$), at $t=const$:
\begin{equation}\label{C_eq}
   C = \int_{\gamma} ds = \int_0^{2\pi} \left|g_{\mu\nu}\frac{dx^{\mu}}{d\phi}\frac{dx^{\nu}}{d\phi}\right|^{1/2}d\phi.
\end{equation}
Then in exponential metric we get the circle-function $C(r)$,
\begin{equation}\label{circ}
   C(r) = \int_0^{2\pi} |g_{33}|^{1/2}(r, \theta = \pi/2)d\phi=2\pi r e^{\varphi(r)}=2\pi r e^{M/r}=2\pi \rho(r),
\end{equation}
where the radial profile
\begin{equation} \label{Rr}
    \rho(r)=r e^{M/r}
\end{equation}
has the same minimal physical radius $\rho_0$  \eqref{minR}, i. e.
\begin{equation} \label{miR}
 \rho_0 = \text{min}\{\rho(r)\} =\rho(M) = eM = 2.71828M,
\end{equation}
and also yields the minimum length of the circle
\begin{equation} \label{minC}
C_0=\text{min}\{C(\rho)\}=\text{min} \{2\pi \rho (r)\}=2\pi \rho_0 =2\pi(eM).
\end{equation}

In other words, in the Papapetrou spacetime at 
$r=M$, there is the smallest permissible
physical circle $C_0=2\pi \rho_0$ \eqref{minC}, determined by
the minimum radius $\rho_0$ \eqref{miR}, and such a procedure turns
out to be equivalent to determining the physical scale of the TWH
throat of the Morris-Thorn type in \eqref{minAr}-\eqref{minR}.

The specified ``bell-shaped'' profile \eqref{Rr} (analog of the ``shape function'' \cite{1988AmJPh..56..395M}) is the spatial part factor of the exponential metric without a singularity at the origin, despite the fact that the concomitant Klein-Gordon equation is satisfied exactly by the singular Newtonian potential.

\begin{figure} 
\begin{minipage}{0.49\linewidth}
\center{\includegraphics[width=0.97 \linewidth]{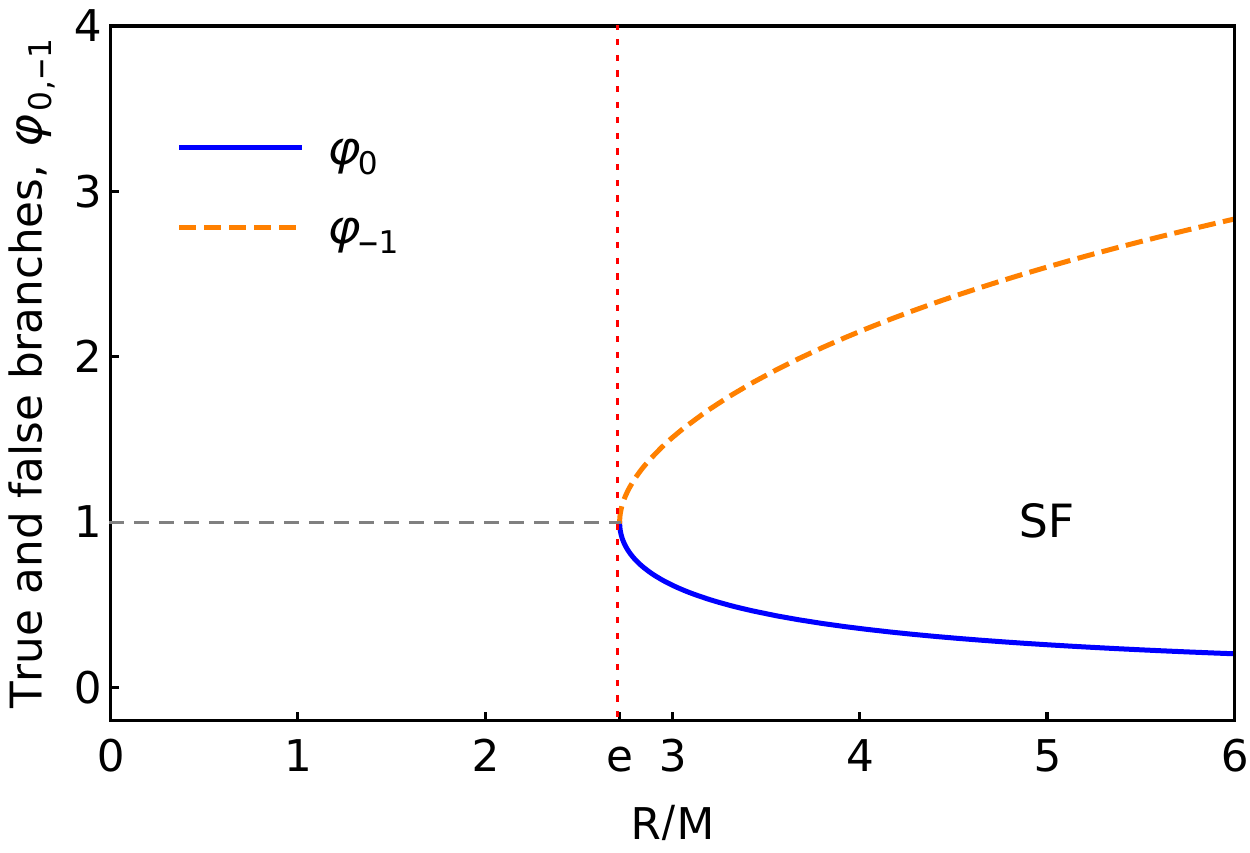}\\ }
\end{minipage}
\hfill 
\begin{minipage}{0.49\linewidth}
\center{\includegraphics[width=0.97\linewidth]{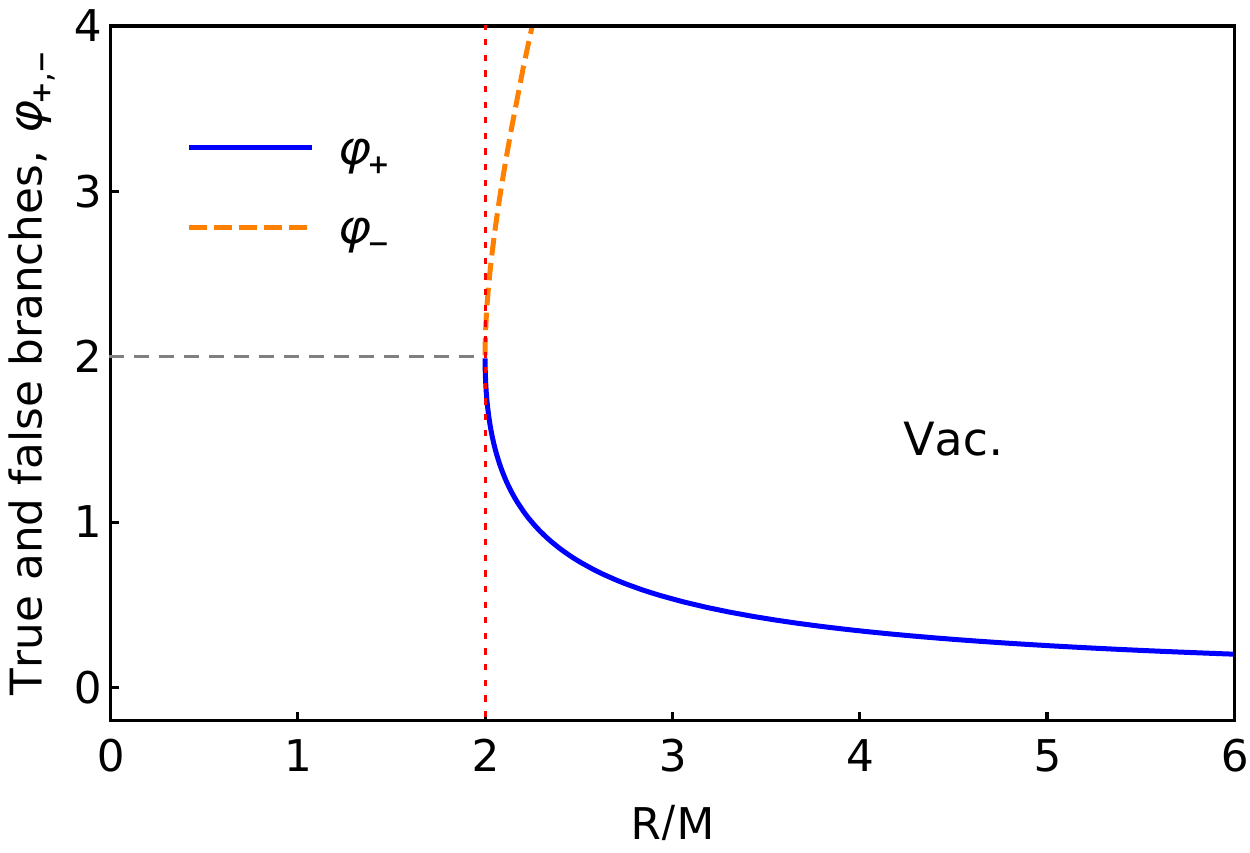}\\ }
\end{minipage}

\caption{ Two-fold representation of the effective Newtonian potential $\varphi(R)$ in curvature coordinates (orange dashed branches) in scalar field background (SF, left) and in vacuum (right) which never covers a region near the origin (marked with red dotted
lines) making it inaccessible for comparison with observations. Only the lower blue branches $\varphi_0 (W_0)$ and $\varphi_+$ might be considered as meaningful.}
\label{fig:potLambert}
\end{figure}

It is interesting that the standard transition in the metric \eqref{Pap} and the Newtonian potential to the curvature coordinates ($t,R,\theta,\phi$) (for details see \cite{2018PhRvD..Visser}, \cite{2024GReGr..56...44M}) analytically coincides with the inversion of the same dependence \eqref{Rr}, namely:
\begin{equation} \label{trasform}
    R = re^{\frac{M}{r}} \quad \Rightarrow \quad r = -M/W\left(-\frac{M}{R}\right), 
\end{equation}
i.e.
\begin{equation}\label{Lamb_metr}
    ds^2(r) \quad \Rightarrow \quad ds^2(R) = e^{-2\varphi(R)} - \frac{dR^2}{\left[1-\varphi(R)\right]^2} - R^2 d\Omega^2,
\end{equation}
\begin{equation} \label{trasform2}
    \varphi(r)=M/r  \quad \Rightarrow \quad \varphi(R) = 
    \begin{cases}
    \varphi_0 = -W_0\left(-\frac{M}{R}\right), \quad R\in[e,\infty]\\
    \varphi_{-1} = -W_{-1}\left(-\frac{M}{R}\right), \quad R\in(e,\infty]
  \end{cases} 
\end{equation}
where $W(x)$ is the two-valued Lambert function. However, such a transition in the potential implies a constraint: at distances smaller than $eM$ information about the scalar field (taking into account both branches of the Lambert function, $W_0$ and $W_{- 1}$ -- see \cite{2018PhRvD..Visser}) is lost, as illustrated by Fig.~\ref{fig:potLambert} (left), where only
the lower branch of the potential graph \eqref{trasform2}, defined by $W_0$, has a physical meaning.

In other words, in the antiscalar mode in the curvature coordinates, there is no mapping $r \rightarrow R$ to the central region $R<eM$.

A similar effect can occur in vacuum \cite{1984PhRvD..29..198G}. If we consider the standard transition $R(r)=r(1+M/2r)^2 \Rightarrow r(R)=\frac{1}{2}\left[R-M\pm (R^2-2MR)^ {1/2}\right]$ from isotropic Schwarzschild coordinates \eqref{Ss} to curvature coordinates:
\begin{equation}\label{}
    ds^2(r) \quad \Rightarrow \quad ds^2(R) = \left(1-2M/R)\right)dt^2 - \left(1-2M/R\right)^{-1}dR^2 - R^2 d\Omega^2,
\end{equation}
then for the effective Newtonian potential we have a two-valued transformation:
\begin{equation} \label{}
    \varphi(r)=M/r  \quad \Rightarrow \quad \varphi(R)=
  \begin{cases}
    \varphi_+ = \frac{2M}{R-M+(R^2-2M)^{1/2}}, \quad R\in[2,\infty]\\
    \varphi_- = \frac{2M}{R-M-(R^2-2M)^{1/2}}, \quad R\in(2,\infty]
  \end{cases}
\end{equation}
where only the $\varphi_+$ branch has physical meaning, and again the region $R<2M$ inside the gravitational radius scale, in curvature coordinates, falls out of consideration, see Fig.~\ref{fig:potLambert} (right). All this restricts the applicability of curvature coordinates in theory and observations, as compared to the isotropic coordinates.

The literature often does not take into account the fact that the variety of existing scalar solutions in other types of coordinates, starting with the Fisher \cite{Fisher1948} and JNW \cite{PhysRevLett.20.878} solutions, are unstable  against collapse and therefore are not interesting from the physical point of view. Only the antiscalar regime leads to a stable and observationally consistent solution, even though arising due to the exotic EMT equation of state. 

Instability of traditional ``scalar'' solutions was first proved by Abe \cite{1988PhRvD..38.1053A} where it has been shown that the perturbation of scalar potential leads to catastrophic (exponential) instability of the initial static solution. Later, this point has been supported and extended \cite{Faraoni21, 2020IJMPD..2950016B}. 
The antiscalar regime might be developed by transfer to a purely imaginary potential \cite{2018PhRvD..98f4050M,2024Univ...10..328G}. In such case the perturbation of this field must be also imaginary, and, as a result, the same linear-perturbation algorithm \cite{1988PhRvD..38.1053A} does not lead to a divergent solution, thereby indicating stability. 

In the the vacuum case all characteristics of the solution are included into the metric. But in scalar background, apart from the metric {\it per se}, we also have scalar field proper defined by the Klein-Gordon equation which, in its turn, is self-consistent with the exponential metric. Hence, the problem of stability might be solved by direct perturbative analysis of the corresponding Klein-Gordon equation as well. This is just what had been done in our earlier work \cite{2018PhRvD..98f4050M}.
Additional justification of the stability of antiscalar regime follows also from thermodynamic stability of scalar field energy-momentum tensor in consideration, as demonstrated in present paper below.

\section{Critical scales for curvature invariants}\label{sec:invars}

Let us continue the analysis of curvature invariants, started in the works \cite{2018PhRvD..Visser} and \cite{2022PDU....3500946T}; we will do it in the context of direct juxtaposition with the Morris-Thorn TWH.

With regard to the search for throat effects in the behavior of algebraic (or ``geometric'') invariants of the Riemann-Christoffel tensor $R_{\alpha\beta\mu\nu}$, we note that the family of these invariants, (see \cite{1997GReGr..29..539Z, Kraniotis_2022}), can be divided into three groups which depend only on the Ricci tensor, or the Weyl tensor, or both. 
\begin{figure*}[ht]
    \centering
    \subfloat[$\sim r^2$]
    {{\includegraphics[width=0.4\linewidth]{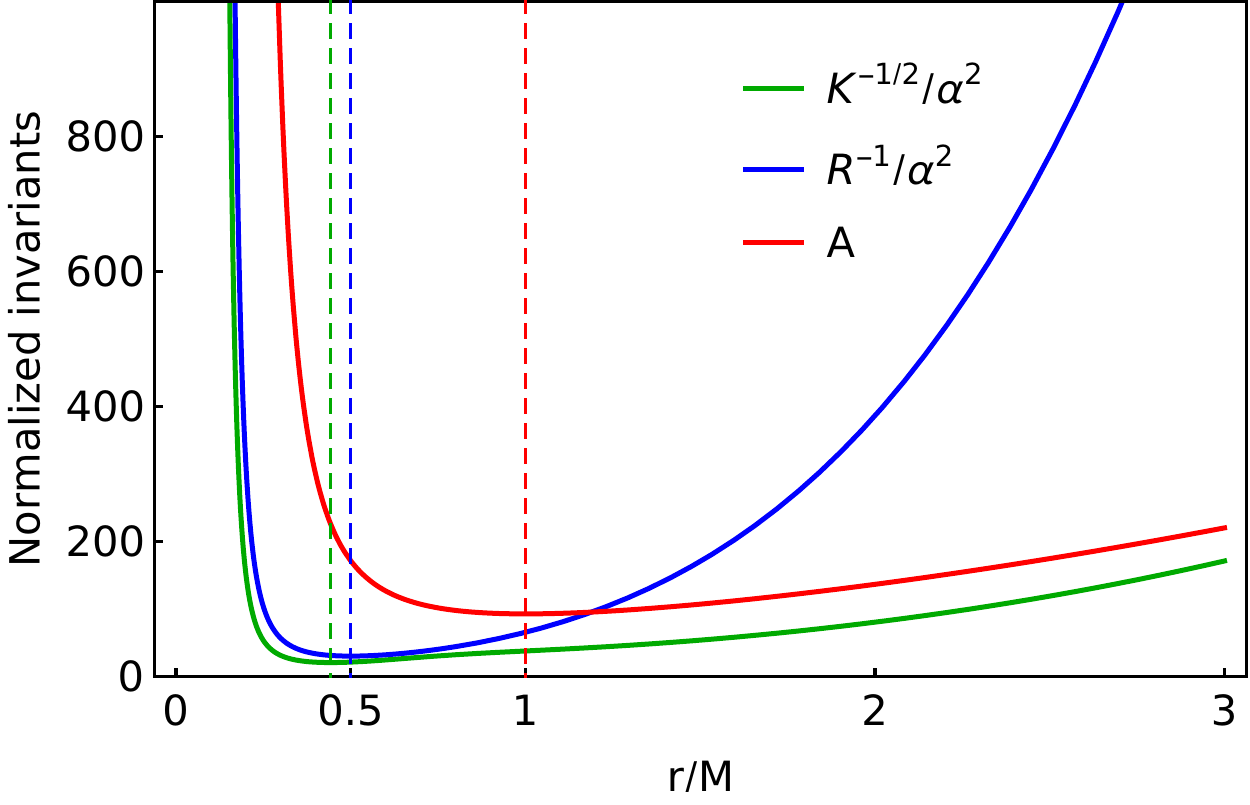} }}%
    \vspace{0mm}
    \subfloat[$\sim r^{-2}$]
    {{\includegraphics[width=0.4\linewidth]{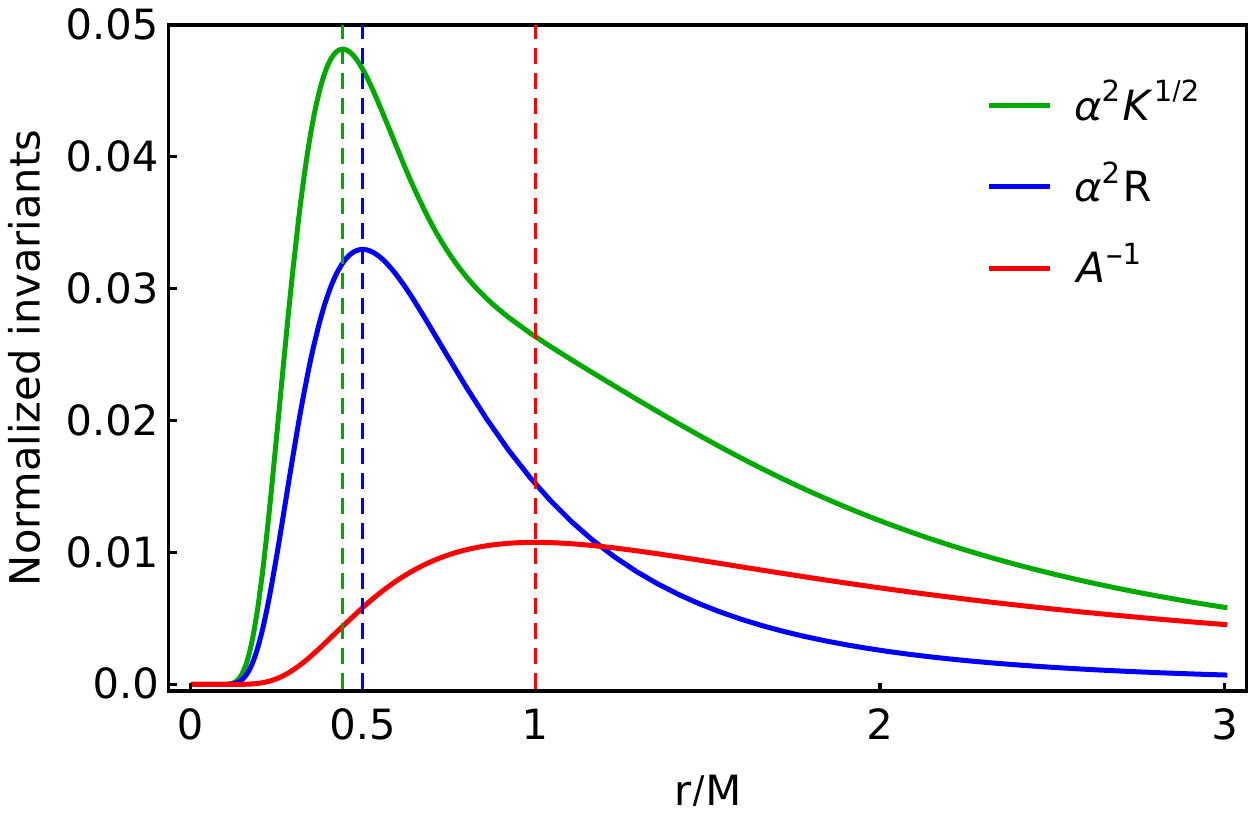} }}%
    \vspace{0mm}
    \subfloat[$\sim r$]
    {{\includegraphics[width=0.4\linewidth]{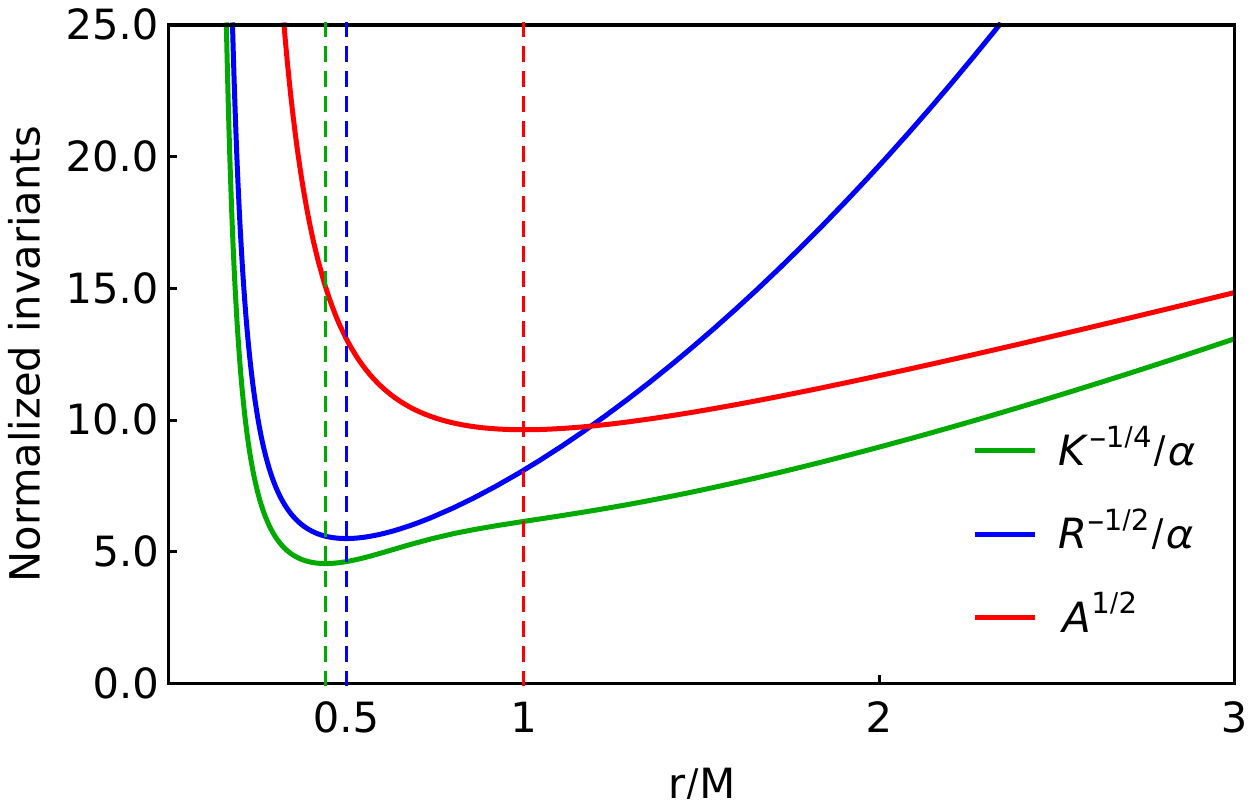} }}%
    \vspace{0mm}
    \subfloat[$\sim r^{-1}$]
    {{\includegraphics[width=0.4\linewidth]{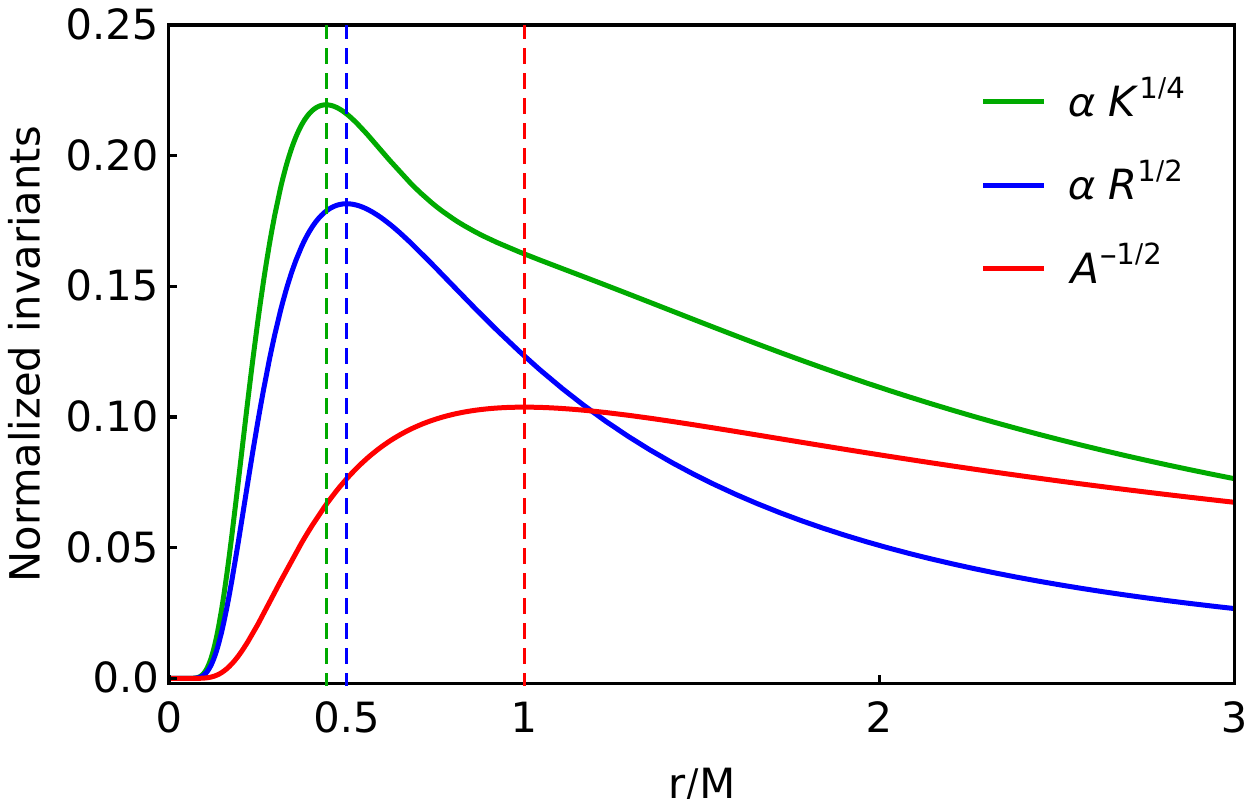} }}%
\caption{ The profiles of the Kretschmann invariants $K$ \eqref{kretch}, Ricci $R$ \eqref{ricci}, and the area $A^{-1}$ of Morris-Thorn TWH resulting from
\eqref{area}. Profiles are transformed to dimensions $[r^ 2]$ (top, left), $[r^{-2}]$ (top, right), $[r]$ (bottom, left), $[r^{-1}]$ (bottom, right) and scaled in accord with \eqref{euler_char}. Vertical dotted lines fix the critical scales at $r_K=0.442 M$ (green), $r_R=M/2$ (blue) and $r_A= M$ (red) which correspond to minima or maxima of corresponding profiles.} \label{fig:invars_Papa_Isotr}
\end{figure*}

To graphically compare the criterion \eqref{area} with the curvature invariants, we scale them according to the Euler characteristic (see \eqref{euler_char} below) and transform those to profiles with the same dimension, such as [$r^2$] and [$r^ {-2}$], or [$r$] and [$r^{-1}$].

This procedure is especially easy to carry out if we notice that in the exponential metric the Kretschmann scalar,
\begin{equation}\label{kretch}
    K=R_{\alpha\beta\mu\nu}R^{\alpha\beta\mu\nu}=R^2\left(7-\frac{16r}{M}+\frac{12r^2}{M^2}\right),
\end{equation}
includes the square of the Ricci scalar $R$ (not to be confused with the
radial curvature coordinate),
\begin{equation}\label{ricci}
    R=R^{\alpha}_{~\alpha}=2\left(\frac{M}{r\rho}\right)^2.
\end{equation}

Then, as shown in Fig.~\ref{fig:invars_Papa_Isotr}, the corresponding [$r^{\pm n}$]-type profiles of the Kretschmann and Ricci invariants have physical extrema at radial distances close to each other $ r_K=0.442 M$ (green) and $r_R=0.5 M$ (blue), and this is completely in accordance with the results of \cite{2022PDU....3500946T}, while in the case of the TWH criterion \eqref{area} this coordinate scale is twice as large, $r_A=M$ (red).

Now, let us consider the standard topological Gauss-Bonnet (GB) invariant $\mathcal{G}$: 
\begin{equation}\label{GB1}
\mathcal{G}= \frac{1}{4}\varepsilon_{\mu \nu \alpha \beta}\,\varepsilon_{\rho \sigma}\,^{\lambda \tau}R^{\alpha\beta}\,_{\lambda \tau}R^{\mu\nu\rho\sigma} = R_{\alpha\beta\mu\nu}R^{\alpha\beta\mu\nu}-4R_{\alpha\beta}R^{\alpha\beta}+R^2,
\end{equation}
with the discriminant tensor $\varepsilon^{\mu \nu \rho \sigma} = \frac{1}{\sqrt{-g}}\epsilon^{\mu \nu \rho \sigma}$ (where $\epsilon_{0123}=-\epsilon^{0123}=1$). 

\begin{figure}
\centering
\includegraphics[width=0.5\linewidth]{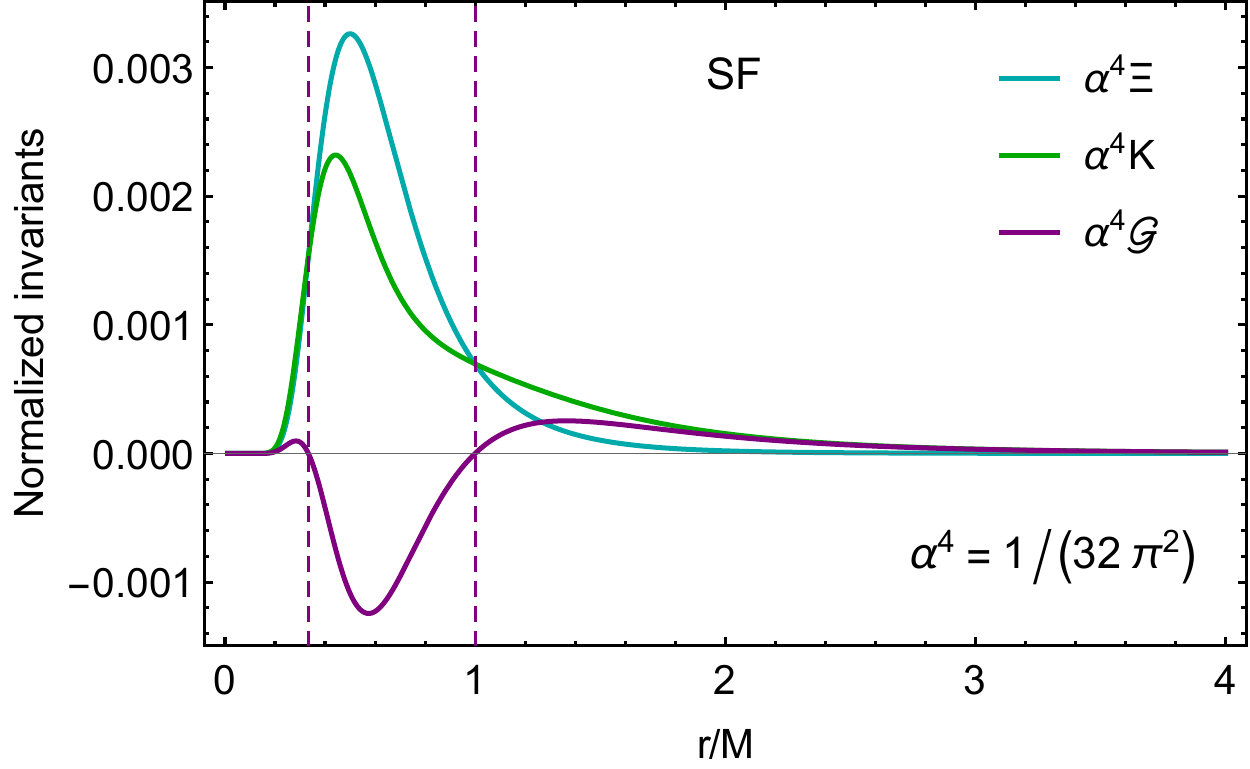}
\caption{The Gauss-Bonnet invariant $\mathcal{G}$ inside the interval $(r=M/3, \,\, r=M)$ becomes negative, which may indicate the concavity of the corresponding generating curve, in contrast to convexity outside this interval, caused by the intersection points of the generating curves $K(r)$ and $\Xi(r)$.} 
\label{fig:GB_fig}
\end{figure}
Inclusion of the invariant $\mathcal{G}$ into the Lagrangian of GR does not change the field equations and their solutions in any way, since it is a 4-divergence. It is important for us that this fundamental scalar has the structure:
\begin{equation}\label{GB}
    \mathcal{G}=K-\Xi ,
\end{equation}
that is, it represents a combination of the Kretschmann invariant $K$ \eqref{kretch} and the so-called thermodynamic invariant (see Section \ref{sec:thermodyn}),
\begin{equation}\label{temp_invar}
    \Xi= 4R_{\alpha\beta}R^{\alpha\beta}-R^2 = 3R^2 = \frac{12M^4}{r^8}e^{-\frac{4M}{r}}.
\end{equation}
The Weyl invariant also has a similar structure (see Appendix \ref{sec:app}).

If spacetime is considered as a compact 4-dimensional Riemannian manifold, the GB invariant is related to the Euler characteristic $\chi$ \cite{2015arXiv150306602B}:
\begin{equation}\label{euler_char}
    \chi=\frac{1}{32\pi^2}\int \mathcal{G} dV = \alpha^4 \int \mathcal{G} dV,
\end{equation}
where the factor $1/(32\pi^2)=\alpha^4$ is used in Fig. \ref{fig:invars_Papa_Isotr} and \ref{fig:GB_fig} for appropriate scaling (normalization) of geometric invariants \footnote{In \cite{2022PDU....3500946T} the authors applied the dimensionless presentation of curvature invariants. Instead of factor $M^4$ of dimension [$r^4$] in \cite{2022PDU....3500946T}, we use the dimensionless factor $\alpha^4$ following from topological relation \eqref{euler_char}. }.

The invariant $\mathcal{G}$ in the form \eqref{GB} has two roots $r=M$ and $r=M/3$ (see Section \ref{sec:thermodyn} and Fig.~\ref{fig:GB_fig}). So, the scale $r=M$, as well as $r=M/3$, is the value at which the topological Gauss-Bonnet invariant changes sign, i. e. at the intersection points of the constitutive graphs $K(r)$ and $\Xi(r)$.

\begin{figure} 
\begin{minipage}{0.49\linewidth}
\center{\includegraphics[width=0.97 \linewidth]{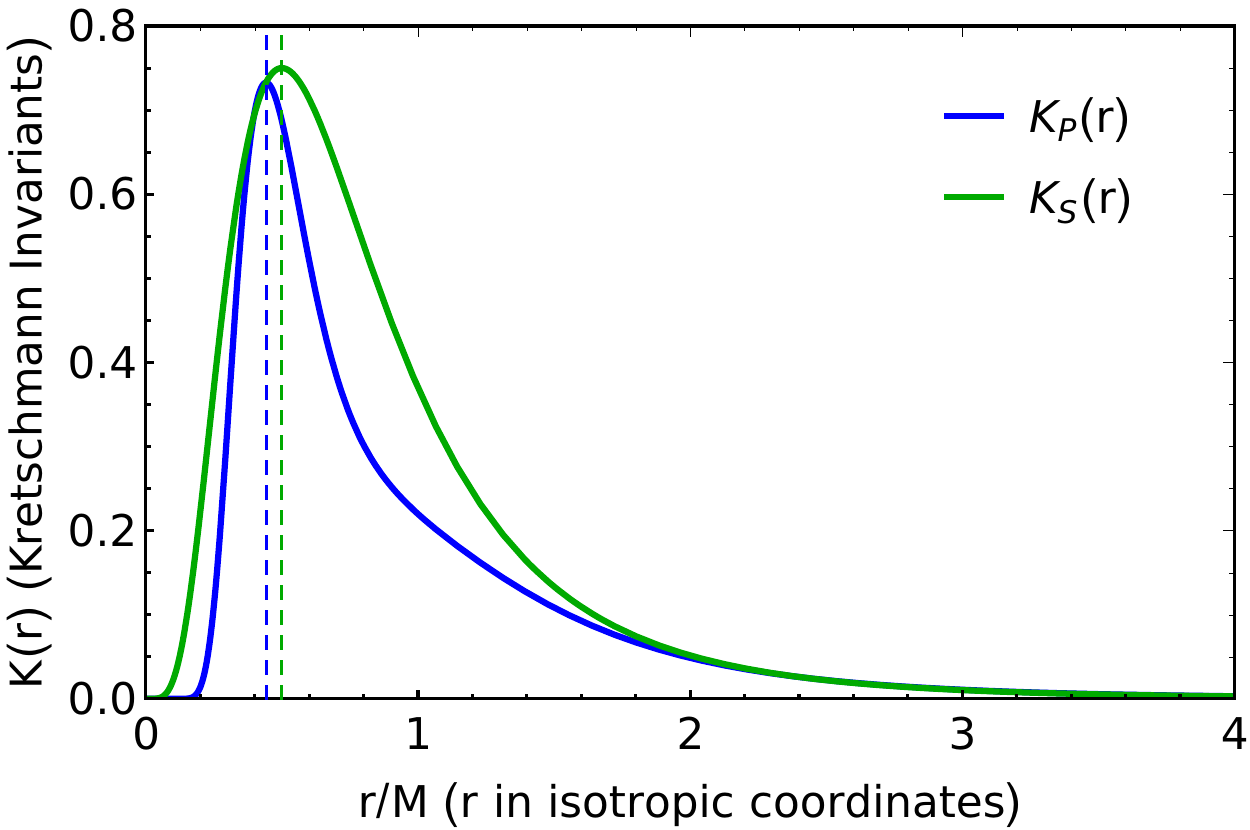}\\ }
\end{minipage}
\hfill 
\begin{minipage}{0.50\linewidth}
\center{\includegraphics[width=0.97\linewidth]{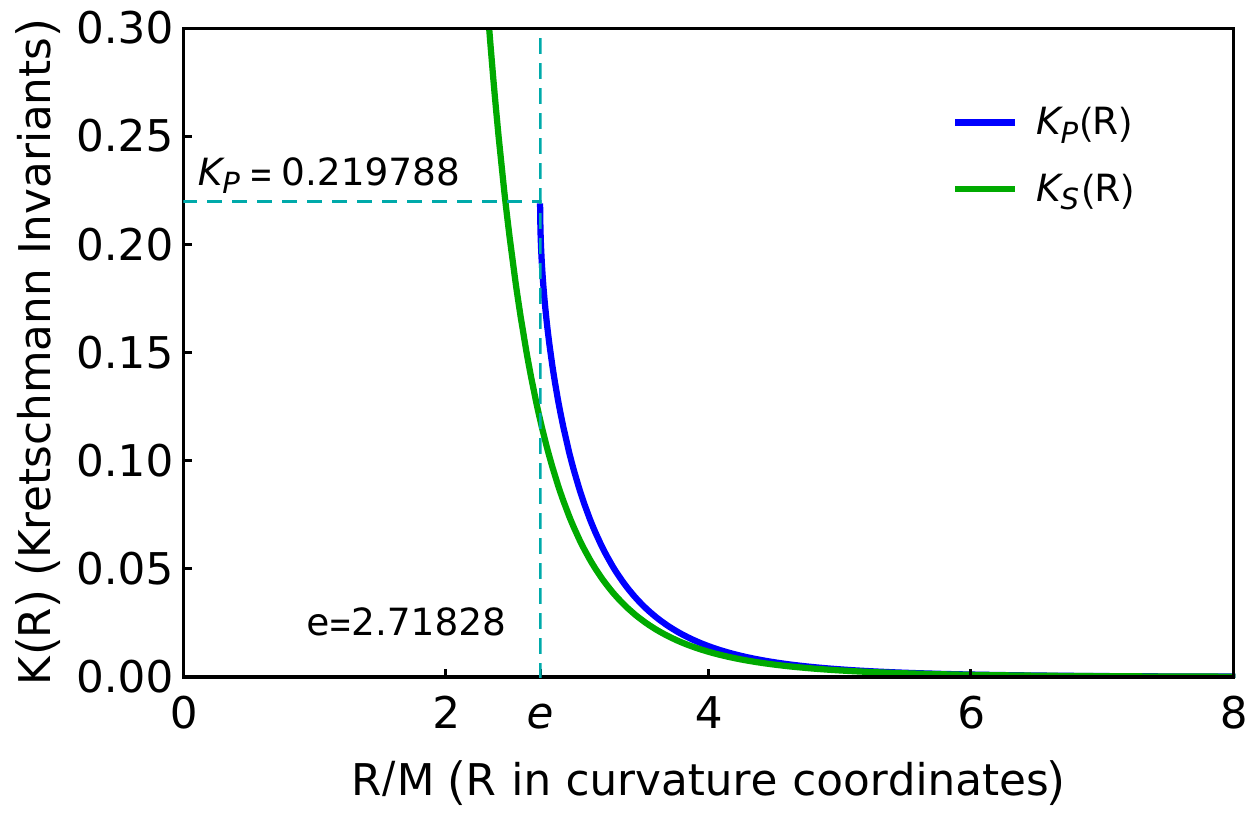}\\ }
\end{minipage}
\caption{\normalfont{\textit{Left}}: The Kretschmann invariants in isotropic coordinates on a scalar background ($K_P$) and in vacuum ($K_S$) are limited, and both at the origin and for $r>2M$ they asymptotically tend to zero, and are maximum at $r/M= 0.442$ (for $K_P(r)$, dotted line, blue) and for $r/M=0.5$ (for $K_S(r)$, dotted line, green).
\normalfont{\textit{Right}}: In curvature coordinates in vacuum, the Kretschmann invariant is singular at the origin (naked singularity). For scalar background $K_P(e)=0.21978$, but information about the invariant $K_P(R)$ in the region $R/M<e=2.71$ is absent in principle, because the mapping \eqref{trasform} does not extend to this region . The Newtonian potential in the indicated coordinates behaves similarly. }
\label{fig:Kretsch_Invar}
\end{figure}

Such behavior is consistent with the interpretation of the exponential metric as a wormhole connecting through its throat the regions of space-time with distinguishable properties. In a sense, that limit $r\rightarrow M$ could also be called ``topological''.

In terms of geometric invariants, a comparison between the scalar background and vacuum is, evidently, possible only for the Kretschmann invariant. 
As shown in Fig.~\ref{fig:Kretsch_Invar}, both metrics reveal a similar non-singular behavior of their curvature invariants in isotropic coordinates not only at the asymptotic distances, but also at the origin.

\section{Singularity of the Keplerian frequency in the Papapetrou metric}
\label{sec:SingKepl}
For other possible effects at $r=M$ we turn to the analysis of the system of equations for geodesics.

Within the framework of the standard algorithm \cite{1999gr.qc....10032W}, based on the expression for the mass shell of a particle of mass $m$,
\begin{equation}
    g_{\alpha \beta}p^{\alpha}p^{\beta}=m^2,
\end{equation}
in arbitrary isotropic coordinates \eqref{isotrop} we have (in the equatorial plane) the radial equation:
\begin{equation}\label{radeq}
\dot{r}^2 = \frac{\tilde{E}^2}{BD} - \frac{1}{D}\left(\frac{\tilde{L}^2}{Dr^2} + 1\right),
\end{equation}
where $\tilde{E}=E/m$, $\tilde{L}=L/m$.
Then in the special case of circular orbits, $\dot{r}=0$, we obtain the relation for the effective potential:
\begin{equation} \label{VeffIsotrop}
V_{eff} = B\left(\frac{\tilde{L}^2}{Dr^2} + 1\right) = \tilde{E}^2.
\end{equation}
Consider two examples.

Example 1. In Schwarzschild (subscript $S$) isotropic coordinates with central mass $M$ we have:
\begin{equation}
V_{eff}^S = \left(\frac{1-\frac{M}{2r}}{1+\frac{M}{2r}}\right)^2\left[\frac{1}{\left(1+\frac{M}{2r}\right)^4}\frac{\tilde{L}^2}{r^2} + 1\right].
\end{equation}
From the condition $\frac{dV_{eff}}{dr} = 0$ we find the angular momentum:
\begin{equation}\label{LS}
\tilde{L}^2(r=r_c) = \frac{4Mr\left(1+\frac{M}{2r}\right)^4}{4-\frac{8M}{r} + \frac{M^2}{r^2}},
\end{equation}
to which corresponds the energy \eqref{VeffIsotrop}:
\begin{equation}\label{ES}
\tilde{E}^2(r_c) = \frac{4\left(1-\frac{M}{2r}\right)^4}{\left(1+\frac{M}{2r}\right)^2\left(4-\frac{8M}{r} + \frac{M^2}{r^2}\right)}.
\end{equation}

\begin{figure}
\centering
\includegraphics[width=0.5\linewidth]{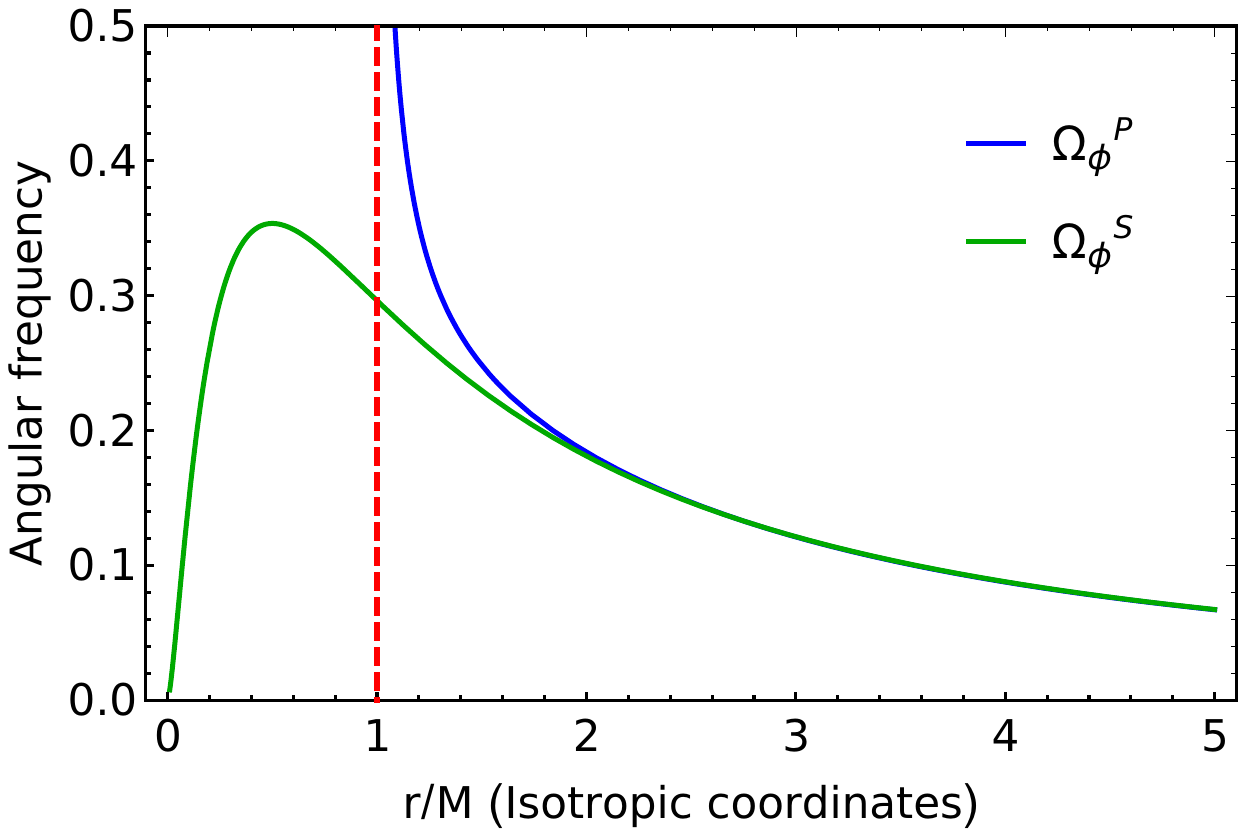}
\caption{ Keplerian frequencies in the isotropic Schwarzschild metric $\Omega_\phi^S$ \eqref{SchwKepler} and in the Papapetrou metric $\Omega_\phi^P$ \eqref{O}, for which the vertical asymptote (dotted line) coincides with the topological limit for the Morris-Thorn type TWH, $r=M$.}
\label{fig:omega_plot}
\end{figure}

\begin{figure} 
\begin{minipage}{0.49\linewidth}
\center{\includegraphics[width=0.98\linewidth]{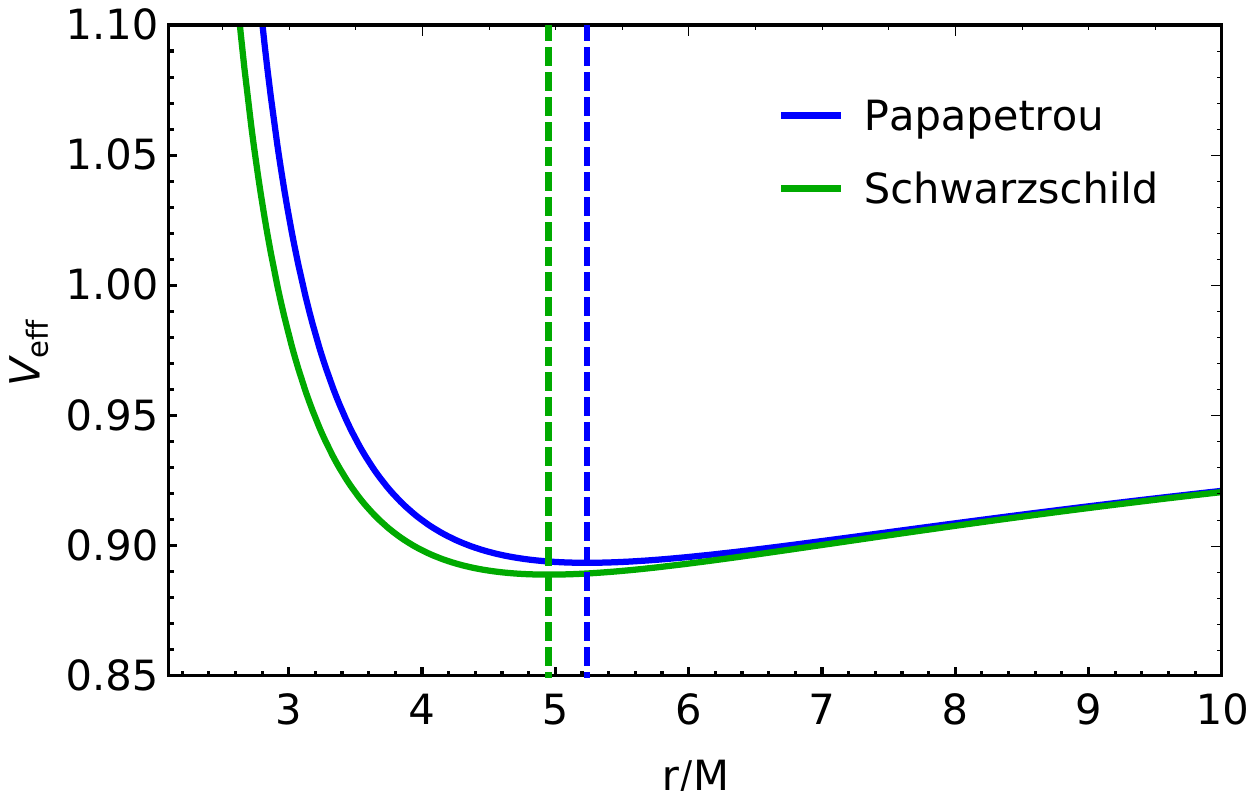}\\ }
\end{minipage}
\hfill 
\begin{minipage}{0.50\linewidth}
\center{\includegraphics[width=0.97\linewidth]{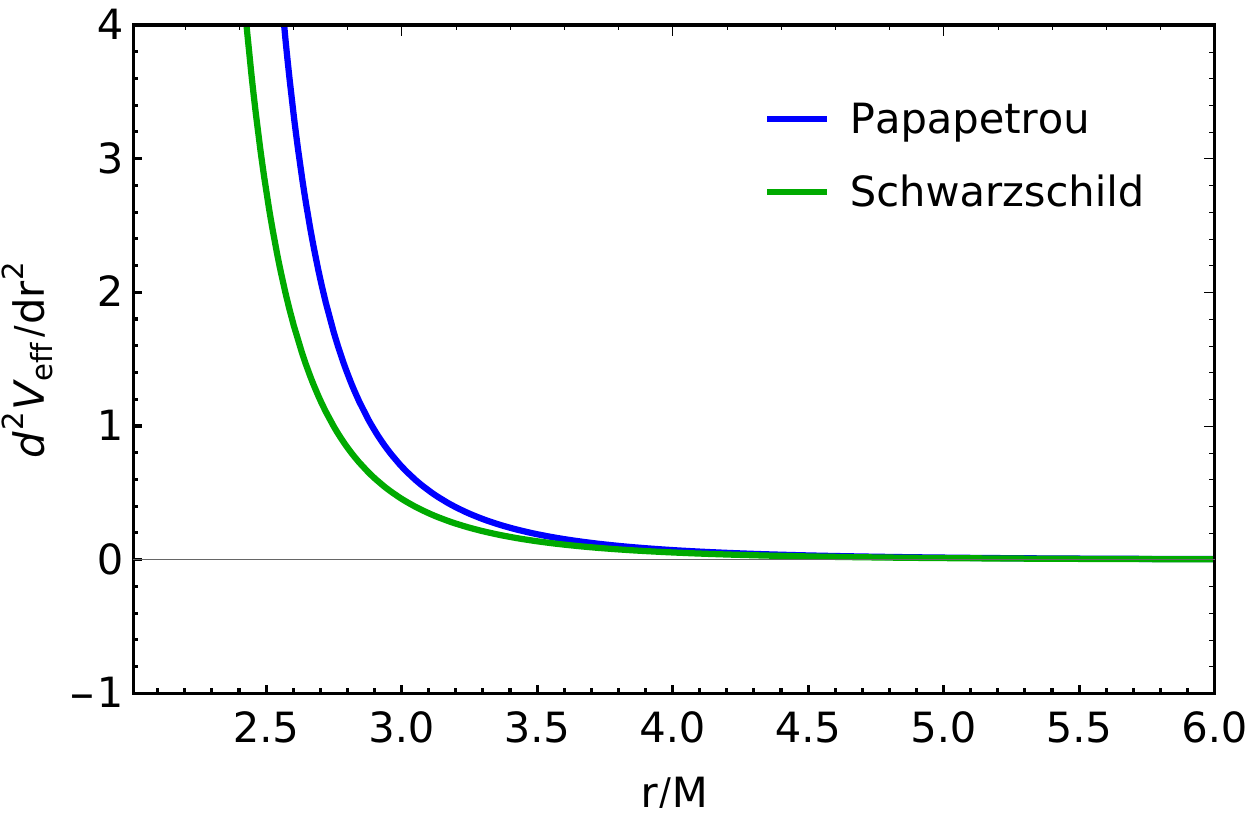}\\ }
\end{minipage}
\caption{Effective potentials $V_{eff}$ the minimums of which correspond to the innermost stable circular orbits with $r_P/M=3+\sqrt{5}$~and $r_S/M =(5+2\sqrt{6})/2$ (left). For $r>2M$, the positive second derivatives of the potentials indicate the stability of circular orbits (right).}
\label{fig:Veff_PS}
\end{figure}

\begin{figure} 
\center
\begin{minipage}{0.49\linewidth}
\center{\includegraphics[width=0.98\linewidth]{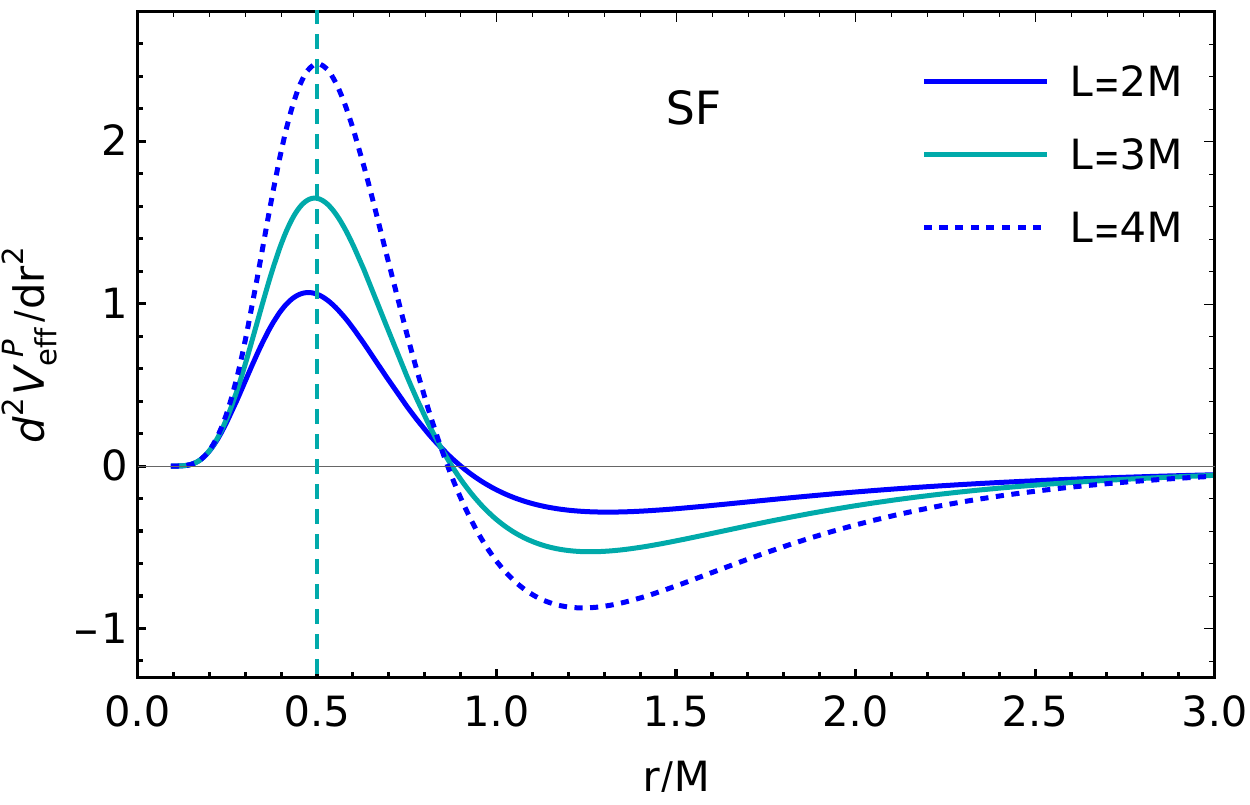}\\ }
\end{minipage}
\caption{ Negative sign in $d^2V_{eff}^P/dr^2$ for arbitrary trajectories as an indicator of instability in contrast to stable circular orbits for $r>2M$ (see Fig. \ref{fig:Veff_PS}). All maxima of stability occur at $r=M/2$ which coincides with the critical Ricci scalar scale.} 
\label{fig:ellipt}
\end{figure}

From here, according to \eqref{Om}, the expression for the Keplerian frequency in vacuum follows (see Fig. \ref{fig:omega_plot}):
\begin{equation}
\label{SchwKepler}
\Omega_\phi^S (r=r_c) =\frac{1}{\left(1+\frac{M}{2r}\right)^3} \sqrt{\frac{M}{r^3}},
\end{equation}
which tends to zero at the origin and at the asymptotics $r\rightarrow \infty$.

Example 2. For circular orbits in the Papapetrou metric (subscript $P$) we have, according to \eqref{VeffIsotrop}, the effective potential \cite{1999gr.qc....10032W}:
\begin{equation} \label{V}
V_{\text{eff}}^P = e^{-\frac{2M}{r}}\left( 1+ e^{-\frac{2M}{r}}\frac{\tilde{L}^2}{r^2}\right)= \tilde{E}^2 \equiv (E/m)^2,
\end{equation}
at the minimum of which (see Fig. \ref{fig:Veff_PS}) the integrals of motion are found - the specific angular momentum and energy of test particles of mass $m$:
\begin{equation} \label{L}
\tilde{L}^2 =(-D \,\,r^2\dot{\phi})^2= e^\frac{2M}{r}\frac{M r}{1-\frac{2M}{r}},
\end{equation}
\begin{equation} \label{E}
\tilde{E}^2 =(B\dot{t})^2 = e^{-\frac{2M}{r}}\frac{1-\frac{M}{r}}{1-\frac{2M}{r}},
\end{equation}
with Keplerian frequency singular at $r=M$ (see Fig.~\ref{fig:omega_plot}):
\begin{equation} \label{O}
\Omega (r\equiv r_{circ})= e^{-\frac{2M}{r}}\sqrt\frac{M}{r^3 \left(1-\frac{M}{r}\right)}.
\end{equation} 
The circular orbits corresponding to \eqref{L} and \eqref{E} are stable for $r>2M$ (see Fig. \ref{fig:Veff_PS}). Within the framework of the system \eqref{L} and \eqref{E} a singularity in \eqref{O} is related to the loss of stability in the Keplerian frequency at $r\leq2M$ through the energy per unit mass in the denominator of \eqref{Om} tending to zero at $r\rightarrow M$.

This is in contrast to instability of elliptical orbits (see Fig. \ref{fig:ellipt}) due to the SO(4) Runge-Lenz effect of precession of the Laplace vector \cite{perelomov1990}, especially in strong fields \cite{2024GReGr..56...44M}. It is noteworthy that the maximal stability scale $r=M/2$ coincides exactly with the extreme point of the Ricci scalar.

So, we find the region of greatest curvature at $r=r_R=0.5M$ for the Ricci scalar and slightly less, at $r=r_K=0.442M$ for the Kretschmann invariant, in agreement with consideration in \cite{2018PhRvD..Visser}, \cite{2022PDU....3500946T}. 

As shown below, the characteristic scale $r=M/2$ also manifests itself within the geometrized thermodynamical approach to the scalar background associated with the Papapetrou metric under consideration.

\section{Geometrized thermodynamics of scalar background}\label{sec:thermodyn}
\subsection{\textit{Geometric and thermodynamic invariants}}
From Einstein’s equations with the perfect fluid EMT,
\begin{equation} \label{EEMT}
    G_{\mu\nu} \equiv R_{\mu\nu} - (1/2)Rg_{\mu\nu}=\varkappa T_{\mu\nu} = \varkappa(\varepsilon u_{\mu}u_{\nu} - Pu_{\mu\nu}),
\end{equation}
where $u_{\mu\nu}\equiv{g_{\mu\nu}-u_{\mu}u_{\nu}}$, follow two main geometric invariants and their thermodynamic equivalents:
\begin{equation}\label{R}
    -R=\varkappa(\varepsilon- 3P),
\end{equation}
\begin{equation}\label{RR}
    R_{\alpha \beta}R^{\alpha \beta} = \varkappa^2(\varepsilon^2+3P^2),
\end{equation}
which might be converted into scalar energy density and pressure:
\begin{equation}
    \varepsilon = (\sqrt{3}/4\varkappa)\Xi^{1/2} - R/4\varkappa,
\end{equation}
\begin{equation}
    P = (1/4\sqrt{3}\varkappa)\Xi^{1/2} + R/4\varkappa,
\end{equation}
with the third invariant $\Xi$, which enters the Gauss-Bonnet scalar $\mathcal{G}=K-\Xi$ as a part completely determined by Einstein’s equations proper, namely ($T\equiv{T ^\alpha}_\alpha)$:
\begin{equation}\label{Xi}
    \Xi \equiv 4R_{\alpha \beta}R^{\alpha \beta}-R^2 = \varkappa^2 (4T_{\alpha \beta}T^{\alpha \beta}-T^2) = 3\varkappa^2(\varepsilon+P)^2.
\end{equation}

\subsection{\textit{$\xi$-scheme of thermodynamics}}
In concordance with kinetic theory of the relativistic gas developed by Synge \cite{1958PhT....11l..56S} the first moment of the distribution function parameterized by temperature, $\Theta=k_BT=\xi^{-1}$, results in the conserved flux density vector $j^{\mu}(\xi)=nu^{\mu}=n\xi^{\mu}/\xi$ with an effective particle number density $n=n(\xi)$, $\xi=\sqrt{\xi^{\alpha}\xi_{\alpha}}$. Then the perfect fluid EMT, as a second conserved  moment, can be identically presented in the form:
\begin{multline}\label{emt}
    {T^{\mu}}_{\nu}=-\frac{\partial j^{\mu}}{\partial \xi^{\nu}} \equiv-\frac{\partial}{\partial \xi^{\nu}}\left(\frac{n \xi^{\mu}}{\xi}\right) =  -\frac{\partial n}{\partial \xi^{\nu}}\frac{\xi^{\mu}}{\xi} - \frac{n}{\xi}\frac{\partial \xi^{\mu}}{\partial \xi^{\nu}} - n\xi^{\mu}\frac{\partial \xi^{-1}}{\partial \xi^{\nu}} = \\  -\frac{\partial n}{\partial \xi}\frac{\xi^{\mu}\xi_{\nu}}{\xi^2} + \frac{n}{\xi}\frac{\xi^{\mu}\xi_{\nu}}{\xi^2} - \frac{n}{\xi}{\delta^{\mu}}_{\nu} = \left(-\frac{\partial n}{\partial \xi}\right)u^{\mu}u_{\nu} - \frac{n}{\xi}{u^{\mu}}_{\nu}, 
\end{multline}
where the energy density $\varepsilon$ and pressure $P$ are expressed as:
\begin{equation}\label{en_pres}
    \varepsilon = -\partial n/\partial \xi, \quad \quad P=n/\xi.
\end{equation}

Next, the first law of thermodynamics for the basic equilibrium systems given by adiabatic Gibbs relation per particle (with an entropy density $s$, a heat flux density $q$, and with zero chemical potential sufficient for our purpose), i. e.
\begin{equation}\label{gibbs_eq}
    dq=\Theta d(s/n)=0=d(\varepsilon/n)+Pd(1/n),
\end{equation}
after substitution of \eqref{en_pres} is transformed exactly into the second-order differential master equation (with primes denoting differentiation w.r.t. $\xi$):
\begin{equation}\label{master_eq}
    nn''+nn'/\xi - (n')^2=0.
\end{equation}
The first integral in \eqref{master_eq} is equation of state with $w=const$:
\begin{equation}\label{EoS}
    -w\partial n/\partial \xi = n/\xi \qquad \Rightarrow  \qquad P=w\varepsilon.
\end{equation}
Now, using some fixed temperature $\Theta_0^{-1}=\xi_{0}$ as a normalizing factor to integrate over dimensionless variable $\hat{\xi}=\xi/{\xi_0}=  \Theta_0/\Theta=\hat{\Theta}^{-1}$, we find that the effective density $n$ is bound to positive constants $n_0>0$ or $C=C(w)=n_0/\Theta_0^{1/w}>0$ (but with arbitrary sign of $w$), as follows:
\begin{equation}\label{eq_n}
    n(w)=n_0\hat{\Theta}^{1/w}=C(w)\Theta^{1/w},
\end{equation}
and, in accord with \eqref{en_pres} and \eqref{gibbs_eq}, we obtain for all thermodynamic quantities as functions of the local temperature $\Theta$:
\begin{equation}\label{epsP}
    \varepsilon(w) = \frac{C(w)}{w}\Theta^{1+\frac{1}{w}}, \qquad P(w) = C(w) \Theta^{1+\frac{1}{w}}, 
\end{equation} 
\begin{equation}  \label{entropy}  
\frac{s(w)}{k_B}=\frac{dP}{d\Theta}=C(w)\left(1+\frac{1}{w}\right)\Theta^{1/w}=n(w)\left(1+\frac{1}{w}\right)=\frac{\varepsilon(w)+P(w)}{\Theta},
\end{equation}
etc. In such equilibrium thermodynamics one can also transfer in Eqs.~\eqref{epsP}, \eqref{entropy} to dimensionless temperature $\hat{\Theta}$ by using \eqref{eq_n}.

This approach is compatible with the kinetic Synge method of moments for the relativistic gas valid for $P\leq\varepsilon/3$ and based on the covariant J{\"u}ttner distribution. But the $\xi$-scheme might also be of use on equal footing for systems with arbitrary equations of state, such as massless scalar field, where the standard kinetic method is not applicable.

In a constant gravitational field (static metric), the dimensionless modulus of the time-like Killing vector is known to be $\hat{\xi} = \sqrt{g_{00}}$. In particular, if
\begin{equation}\label{hateps}
    \hat{\xi} \equiv{\frac{\xi}{\xi_0}} = \frac{\Theta_0}{\Theta} = \sqrt{g_{00}},
\end{equation} 
then $\Theta_0=\Theta\sqrt{g_{00}}$ might also correspond to the temperature, which remains the same during the movement of a given fluid element \cite{landaustat} in 4-dimensional spacetime. 
Here, all the results for the peculiar systems described by \eqref{eq_n}-\eqref{entropy} are represented only in terms of the local temperature $\Theta(r)$ measured at the given point.

\subsection{\textit{Scalar background and its temperature}}
Now, let's consider the thermodynamics of an equilibrium scalar field in terms of geometric invariants. 

Let us define the 4-velocity $u_{\mu}=\varphi_{\mu}/\sqrt{\varphi_{\alpha}\varphi^{\alpha}}$. 
Substitute it into Einstein's equations \eqref{EEMT}. Then, matching with \eqref{EinEq} we get $P=\frac{1}{2}\varphi_{\alpha}\varphi^{\alpha}$. So, for the time-like gradient $\varphi_{\mu} \equiv \varphi_{,\mu }$ one gets the rigid equation of state $P=\varepsilon,\, w=1$ which fixes the integration constant, $C(w)=C(1)\equiv C_1$, and so, from \eqref{eq_n}-\eqref{entropy} one gets in scalar background:
\begin{equation}\label{entr}
    n=C_1\Theta, \quad \varepsilon = P = C_1\Theta^2, \quad \frac{s}{k_B}=\frac{dP}{d\Theta}=2C_1\Theta=2n=\frac{2\varepsilon}{\Theta}.
\end{equation}
In exponential metric, invariants \eqref{R}, \eqref{RR}, \eqref{Xi} are reduced to the Ricci scalar:
\begin{equation} \label{R0}
R\,\, \dot{=}\,\, 2\varkappa \varepsilon,
\quad
    R_{\alpha \beta}R^{\alpha \beta} = 4\varkappa^2 \varepsilon^2 = R^2,
\quad
    \Xi \equiv 4R_{\alpha \beta}R^{\alpha \beta}-R^2 = 12\varkappa^2 \varepsilon^2 = 3R^2,
\end{equation}
and, therefore, reach the maximum value at the same radial distance $r=M/2$.

Next, from \eqref{R0}, \eqref{kretch} and \eqref{ricci} in the same metric \eqref{Pap} we find the intersection points of the invariants $K$ and $\Xi$ as the roots $r=M/3 $ and $r=M$ of the equation $$ \mathcal{G}(r)=K(r)-\Xi(r)=\frac{16e^{-4M/r}M^2(M^2-4Mr +3r^2)}{r^8} =0,$$ as already noted in Section \ref{sec:invars}. The negative sign of the topological invariant $\mathcal{G}(r)$ in the interval between $r=M/3$ and $r =M$ may indicate that in this region the character of the curvature of space has changed from convex to concave.

As a consequence of state with $w=1$, according to \eqref{R0} and \eqref{entr}, one obtains that the Ricci scalar in exponential metric is proportional to the square of the temperature:
\begin{equation} \label{RRR}
    R \equiv R(r)= 2 \frac{ G^2 M^2}{c^4 r^4} \exp\left(\frac{-2GM}{c^2r}\right) \equiv 2 \frac{ G^2 M^2}{c^4 r^4}g_{00} \,\, \dot{=}\,\, 2\varkappa C_1 \Theta^2 .
\end{equation}
Hence, in accord with \eqref{entr}--\eqref{R0}, the physical characteristics of scalar background (e. g., temperature, energy density or entropy) might be geometrized within the framework of GR with the next relationship to the Ricci scalar:
\begin{equation} \label{invTT}
  \Theta = \frac{1}{\sqrt{2\varkappa C_1}}\sqrt{R(r)},\quad 
  \epsilon=\frac{\sqrt{R_{\alpha \beta}R^{\alpha \beta}}}{2\varkappa} = \frac{R}{2\varkappa}, \quad \frac{s}{k_B}=2C_1\Theta = \sqrt{\frac{2C_1 R(r)}{\varkappa}},
\end{equation}
which, with constant $\varkappa\equiv8\pi G/c^4$, for the local temperature of scalar field in the Papapetrou spacetime yields relation (reminiscent of but non-equivalent to \eqref{hateps}):
\begin{equation} \label{locT}
   \Theta(r)=\frac{1}{\sqrt{\varkappa C_1}}\frac{GM}{c^2 r^2}\exp\left(-\frac{GM}{c^2 r}\right)=\Theta^*(r)\sqrt{g_{00}}.
\end{equation}
In particular, $\Theta=\Theta(r)= k_B T(r)$ has physical maximum at $\bar{r}=GM/2c^2$ which, granted $\sqrt{g_{00}(\bar{r})}=e^{-2},\, e=2.718...$, and up to arbitrary $C_1$ will be
\begin{equation} \label{maxT}
   \Theta(\bar{r})=\frac{4e^{-2}}{\sqrt{\varkappa C_1}}\frac{c^2}{G}\frac{1}{M}.
\end{equation}
This is the ``hottest'' region in scalar background, which, note, does not coincide with the coordinate distance $r=GM/c^2$ in the approach based on the Morris-Thorn TWH. So, invariant temperature and other physical quantities on the scale $r=M$ (with $G=c=1$) do not have extrema. 

As for the constant $C_1$, in particular, it could be estimated by juxtaposition of temperature \eqref{locT} with the known Hawking black hole temperature calculated in the curvature coordinates (in usual units):
\begin{equation} \label{rg}
        \Theta_{\text{BH}}(R_g) \equiv k_B T_{BH}(R_g) = \frac{\hbar c}{4\pi}\frac{1}{R_g} \equiv  \frac{\hbar c^3}{8\pi G}\frac{1}{M} \,.
\end{equation}

However, to be correct, the standard gravitational radius $R_g=2GM/c^2$ should be expressed through the Schwarzschild ``isotropic'' horizon $r_g=\bar{r}=GM/2c^2$ following from $g_{00}(r_g)=0$. Numerically, we get the same expression as for the extremal value $\bar{r}$ within the relation \eqref{locT} in the isotropic Papapetrou coordinates without horizons. So, with the transfer $R_g\rightarrow\bar{r}$ we obtain:

\begin{equation}  \label{isoT}
     T_{\text{BH}}(\bar{r})= \frac{\hbar c}{4\pi k_B}\frac{1}{\bar{r}\left(1+\frac{GM}{2c^2 \bar{r}} \right)^2}=\frac{\hbar c^3}{8\pi k_B G}\frac{1}{M}\approx 10^{-7}\frac{M_{\odot}}{M}K.
\end{equation}

Then the integration constant $C=C_1$ can be found by comparing \eqref{isoT} with \eqref{maxT} on the scale $\bar{r}$: $\Theta(\bar{r})=\Theta_{\text{BH}}(\bar{r})$. In accord with \eqref{locT} this yields in isotropic coordinates:
\begin{equation}   \label{C1}
  \frac{1}{\sqrt{\varkappa C_1}}= \frac{\hbar c}{32\pi}\frac{1}{\sqrt{g_{00}(\bar{r})}} = \frac{e^2\hbar c}{32\pi}\, \Rightarrow\, \Theta(r)=k_B T(r)=\frac{e^2}{32\pi}\frac{\hbar GM}{c r^2}\exp\left(-\frac{GM}{c^2 r}\right).
\end{equation}

This estimation differs from our previous result \cite{2018PhRvD..98f4050M} by a numerical factor of the order of unity due to transfer to isotropic frame and taking into account the contribution of the factor $\sqrt{g_{00}(\bar{r})}$, while in \cite{2018PhRvD..98f4050M} we used the approximate condition $\Theta^*(r=R_g)=\Theta_{\text{BH}}(R_g)$ expressed in curvature coordinates. In both cases the coincidence with vacuum thermodynamics (at the same profile point) is guaranteed by construction.
\begin{figure}[ht] 
\center
\begin{minipage}{0.49\linewidth}
\center{\includegraphics[width=0.98\linewidth]{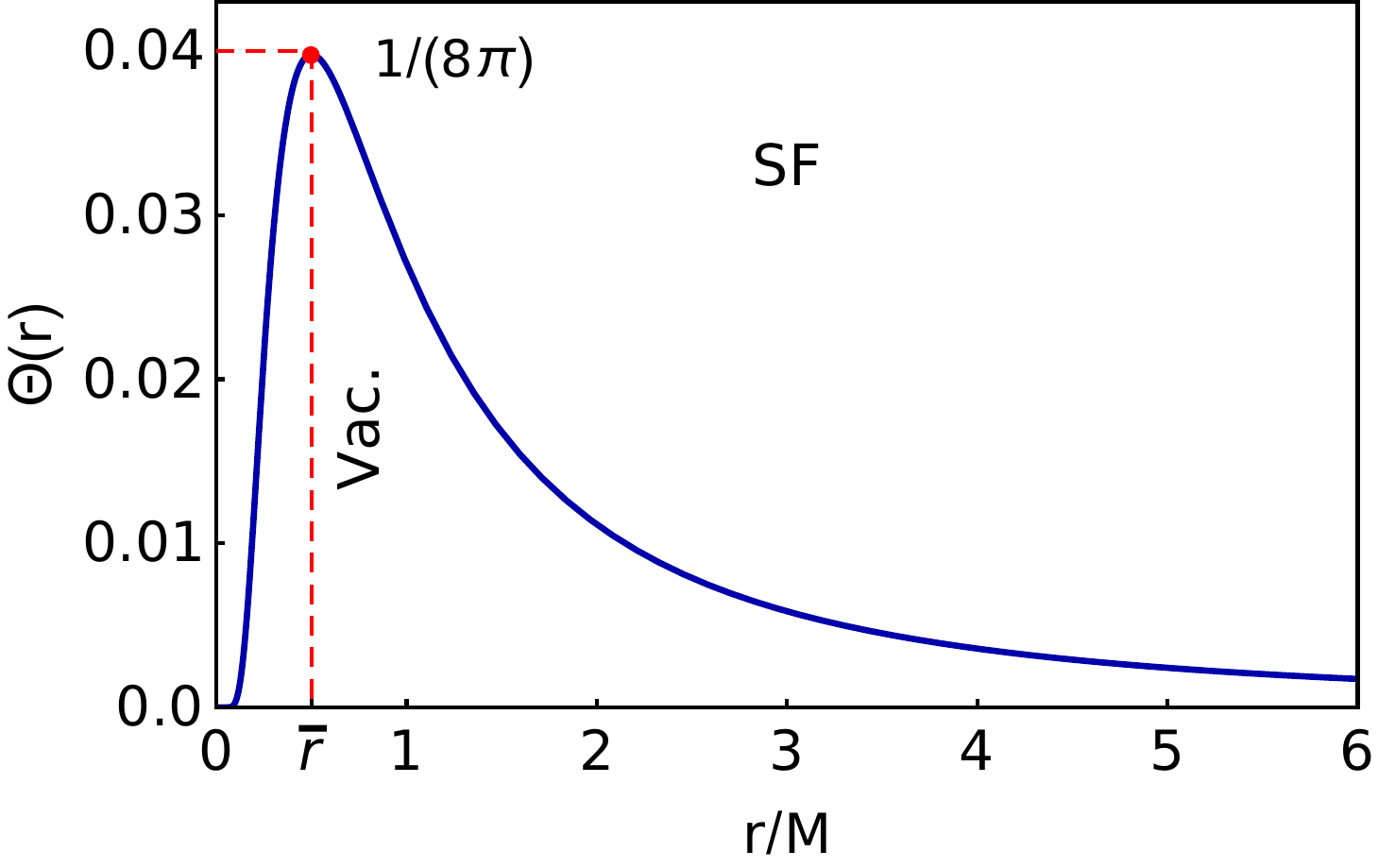}\\ }
\end{minipage}
\caption{ The distribution of temperature of scalar background in exponential metric in units $\hbar=c=G=k_B=1$. Here, the extremal point is set to be equal to the BH-temperature (dashed line) at $\bar{r}=M/2$.}
\label{fig:temperatureProfile}
\end{figure}

The radial profile for the scalar background temperature \eqref{C1} is shown in Fig.~\ref{fig:temperatureProfile}. It defines the distribution of the ambient scalar field temperature around the central mass (scalar charge) $M$
with the critical scale $\bar{r}=GM/2c^2$ (corresponding to the gravitational radius $R_g$ in the curvature coordinates). The distributions of other thermodynamic characteristics depending on the temperature might be calculated by \eqref{entr}--\eqref{invTT} and at the same critical scale they reduce to the known results of equilibrium thermodynamics of black holes connected to horizon (see also \cite{2018PhRvD..98f4050M}). 

In such a gauge, this profile might be interpreted as a spreading of the fixed horizon temperature, which appears to be an expected physical effect conditioned by the scalar background. Analogously, exponential metric leads to the distribution of other static equilibrium thermodynamic quantities. But due to absence of any horizons, such dynamical effects as the Hawking radiation do not arise in the Papapetrou spacetime where any `in- and out-' virtual quantum particles and antiparticles arising close to some spherical surface would immediately be canceled out without generating the outgoing quantum thermal radiation. 

\section{Electrostatic approach to scalar background and origin of antiscalar regime}
\label{sec:electostatic} 

It is difficult to overestimate the meaning of identifiable scalar fields. As a first step in this regard, we will consider the relation between the Einstein-Maxwell and Einstein-scalar equations, with resulting presentation in the Papapetrou spacetime. 

From the minimal Einstein-scalar equations in antiscalar regime,
\begin{equation}
G_{\mu\nu} = -\varkappa T_{\mu\nu}^{SF} = -\varkappa \frac{1}{4\pi}  \left( \varphi_{\mu} \varphi_{\nu} - \frac{1}{2} g_{\mu\nu} \varphi_{\alpha} \varphi^{\alpha} \right), \quad \varphi_{\mu}\equiv\varphi_{,\mu},
\label{A1}
\end{equation}
in case of static spherical symmetry (a point-mass $M$ source), $\varphi = \varphi(r)$, one gets
\begin{equation}    
G_{\mu \nu} = - \varkappa \, \text{diag} \{  T_{00}^{SF}, T_{11}^{SF},
T_{22}^{SF}, T_{33}^{SF}\} \quad \text{(antiscalar regime)},
\end{equation}
with the following local values of components and trace of EMT:
\begin{equation}
T_{00}^{SF} =- \frac{1}{8\pi} (\varphi_\alpha\varphi^\alpha), \quad T_{ij}^{SF} = \frac{1}{4\pi}
\left( \varphi_i \varphi_j - \frac{1}{2} g_{ij} \varphi_{\alpha} \varphi^{\alpha}  \right)\,\Rightarrow\,
^{SF}{T_{\alpha}}^{\alpha} = -\frac{1}{4\pi} \left( \varphi_{\alpha}
\varphi^{\alpha}  \right).
\label{TSF}
\end{equation}

On the other hand, the standard Einstein-Maxwell equations, 
\begin{equation}
G_{\mu\nu} = \varkappa T_{\mu\nu}^{EM} = \varkappa \frac{1}{4\pi}
\left( -F_{\mu \alpha} {F_{\nu}}^{\alpha} + \frac{1}{4} F_{\alpha
	\beta } F^{\alpha \beta } g_{\mu \nu}\right),
\label{A2}
\end{equation}
where $F_{\mu\nu} = \partial_{\mu} A_{\nu} - \partial_{\nu} A_{\mu}$, in spherically symmetric electrostatic case $$A_\mu \to
\{\Phi(r), 0, 0, 0 \},$$ have a similar matrix structure,
\begin{equation}
G_{\mu \nu} = \varkappa \, \text{diag} \{  T_{00}^{EM}, T_{11}^{EM},
T_{22}^{EM}, T_{33}^{EM}\} \quad \text{(standard regime)},
\end{equation}
with components depending only on  $\Phi_r \equiv \Phi_{,r}$ of $\Phi_{\mu} = \{0, \Phi_r, 0, 0\}$:
\begin{equation}
T_{00}^{EM} =- \frac{1}{8\pi} (\Phi_\alpha\Phi^\alpha), \quad T_{ij}^{EM} = -\frac{1}{4\pi}
\left( \Phi_i \Phi_j - \frac{1}{2} g_{ij} \Phi_{\alpha} \Phi^{\alpha}  \right)\,\Rightarrow\,
^{EM}{T_{\alpha}}^{\alpha} = 0.
\label{TEM}
\end{equation}

We see that in the static limit, the $T_{ij}^{EM}$ components of the Einstein-Maxwell equations \eqref{A2} coincide with the corresponding components of the Einstein-scalar equations \eqref{A1} exactly in antiscalar form, $-T_{ij}^{SF}$, with simultaneous equality of the $(0, 0)$ values for the energy densities in \eqref{TEM} and \eqref{TSF}. This is a prerequisite for the manifestation of a certain combination of charged electric fields as an effective neutral scalar field with non-zero EMT trace.

Specifically, the quasi-static-type field $\Phi(r)$ always implies $\Phi_+(r)$ or $\Phi_-(r)$ as constituents proportional to electric charges, i. e. $\Phi_+ \sim q_+$ or $\Phi_- \sim q_-$, which typically exist pairwise localized, being asymptotically compensated in the electrodynamic sense, but with residual potentials being able to participate in gravitational interactions. Such picture complies with neutral, on average, universe, so, one might suggest for all pairwise localized in space quasi-electrostatic fields the following ansatz leading at long-range, on average, to neutral electro-vacuum background: 
\begin{equation}
\varphi(r) = \langle
\frac{\alpha}{2}( |\Phi_+(r)| + |\Phi_-(r)| )\rangle=\frac{M}{r}. 
\label{electricansatz}
\end{equation}
This relationship contains the dimensional factor $\alpha={M}/{q}$,
in conformity with the exponential Papapetrou metric with scalar charges reduced to the masses (in units $G=c=1$, \cite{2020FoPh...50.1346M}) and with the Newtonian-type potential as a non-asymptotic but rigorous solution \cite{mm19} of the same form as the Coulomb field up to the factor $\alpha$. 

From the thermodynamic point of view, such scalar background medium possesses the stiff equation of state, $\varepsilon=p$, and this medium should obey a thermodynamic stability which requires the second-order derivative of the thermodynamic energy $E$ (intensive value) with respect to volume $V$ (extensive value) to be positive. 
Indeed, going from integral quantities $E, V$ into densities $\varepsilon, n$, thus replacing the volume derivatives, $\partial/\partial V \, \rightarrow \, \partial/\partial (1/n) = -n^2 \partial/\partial n$, and taking into account the standard laws for pressure,  $p=w \varepsilon$ and  $p = n \partial \varepsilon / \partial n - \varepsilon,$ we get:
\begin{equation}
	\frac{\partial^2 E}{\partial V^2} = -\frac{\partial p}{\partial V} > 0 \quad \Rightarrow \quad n^2 \frac{\partial \varepsilon}{\partial n }w>0,
\end{equation}
i. e. the stability condition for systems with the state parameter $w$ becomes
\begin{equation}
\frac{\partial \varepsilon}{ \partial n} = (1+w)\frac{\varepsilon}{n} \quad \Rightarrow \quad 
w(w+1) n \varepsilon >0.
	\label{condnes}
\end{equation}
So, for $\varepsilon > 0$, it follows: $w(w+1)>0$. Comparing the standard scalar field EMT \eqref{A1} with the perfect fluid EMT, ${T}_{\mu\nu}^\text{pf} = (\varepsilon + p)u_\mu u_\nu - p {g}_{\mu\nu}$,
for $u_\mu = \phi_\mu /\sqrt{\phi_\alpha \phi^\alpha}$ we get for time-like gradient $\phi_\mu$ ($\phi_\alpha \phi^\alpha>0$, $u_\alpha u^\alpha = 1$) the state $w = 1$ which is stable. 

Thus, for ambient scalar field in stiff state $p=\varepsilon$ (suggesting only transverse modes), the permanent ``sound'' speed $(d p/d\varepsilon)^{1/2}=v/c=1$ in a medium comprising a neutral superposition of quasistatic electric fields will be identical to ordinary speed of light. 

This completes the presentation of the massless scalar background as, on average, a neutral aggregate of the primordial electrovacuum states.

 \section{Conclusion and outlook}
\label{sec:conc}

We have revealed only two features related directly to the TWH throat scale, $r=M$. These are the topological Gauss-Bonnet invariant which at the given distance changes sign, and the Keplerian frequency, which in the domain of unstable circular geodesics at $r\rightarrow M$ converges to singularity due to proper energy of test particles simultaneously converging to zero, in accord with condition \eqref{Om}. 

At the same time, the exponential metric is characterized by other extremal points. Indeed, the Ricci scalar, as well as other invariants issuing directly from the field equations of general relativity, all have extrema at the same point, $r=M/2$, which (unlike the TWH throat) entails a number of physical effects.  In particular, at that scale we get the maximal temperature generated by the exponential metric. The complete temperature profile gives thermodynamic description of scalar background in space and yields, at the scale of horizon in the Schwarzschild isotropic coordinates, the known equilibrium black hole thermodynamics relations. 
    
It is noteworthy that the scale $r=M/2$ is manifested as more physically relevant than $r=M$, which appears to be an artifact of special geometric definition of a throat, rigidly restricted by the spherical symmetry in question.

As for the actual problem of identification of the massless scalar background, we have shown that in static limit the standard Einstein-Maxwell equations reveal the antiscalar regime for electric fields. So, the ambient scalar field might, in principle, be comprehended as electrically neutral superposition of primordial quasistatic fields and, in its turn, acts as a stiff and thermodynamically stable medium.

At the cosmological scales, a new physical situation might arise \cite{2015IJMPD..2444025M,mm19}, if the vanishingly small but non-zero masses of the background scalar field mediators are taken into account, which proves to be  tightly connected with the cosmological constant value and thereby might be naturally related to dark energy.

\appendix

\section{The structure of conformal invariants}\label{sec:app}
The structure of invariants that depend on the Ricci tensor shows that they all have physical extrema only at the coordinate distance $r=M/2$. We are talking about such invariants
(using the numbering from \cite{Kraniotis_2022}) as
\begin{eqnarray}
I_5&=&R^{\alpha}_{\alpha}=R,\\
I_6&=&R_{\alpha \beta}R^{\alpha \beta},\\
I_7&=&R_{\mu}^{\nu}R_{\nu}^{\lambda}R_{\lambda}^{\mu},\\
I_8&=&R_{\mu}^{\nu}R_{\nu}^{\lambda}R_{\lambda}^{\rho}R_{\rho}^{\mu}.
\end{eqnarray}
\begin{figure}[ht] 
\centering
\includegraphics[width=0.7\linewidth]{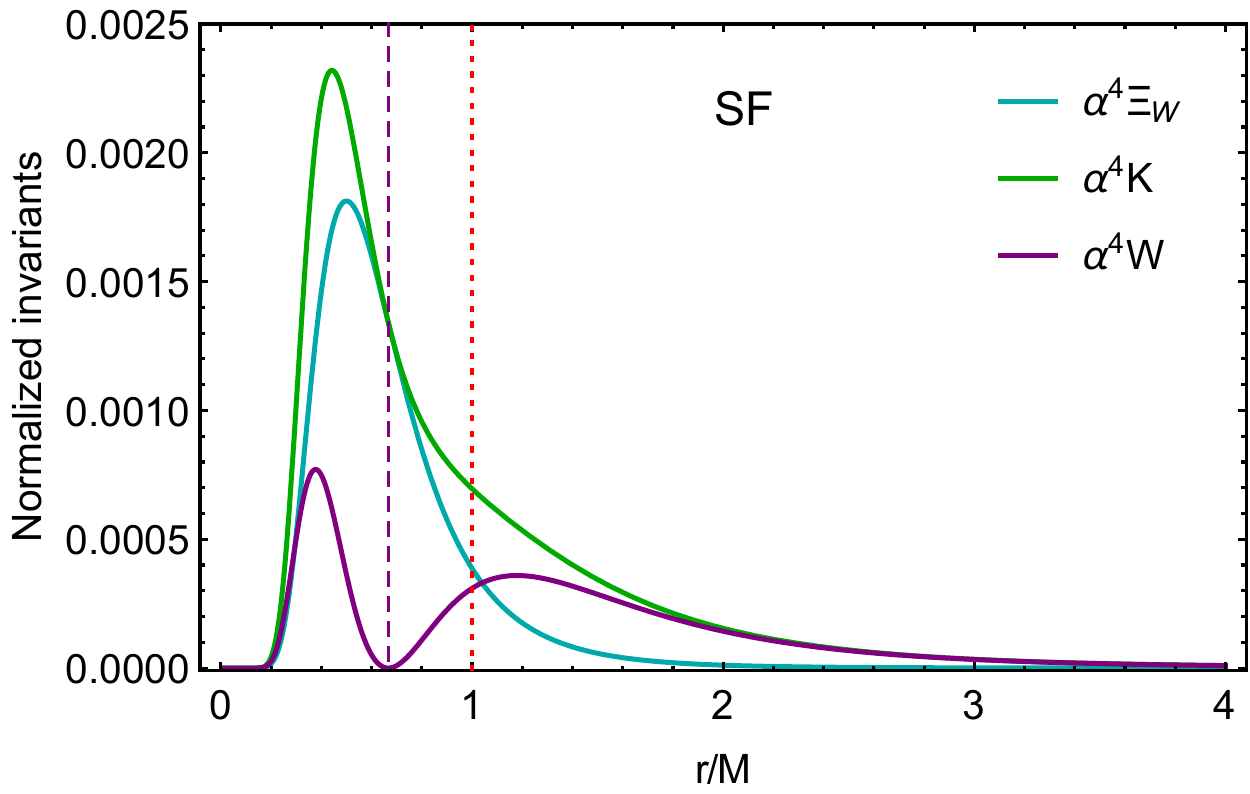}
\caption{The Weyl invariant vanishes at the point $r=2M/3$, but does not change sign anywhere, since the generating profiles of the invariants $K$ and $\Xi_W$ touch each other at the designated point, but do not intersect each other, as in the case with $\mathcal{G}$. In this case, none of the extrema is associated with the coordinate scale $r=M$ (dotted, red).} 
\label{fig:Weyl_fig}
\end{figure}
\begin{figure}[ht] 
\centering
\includegraphics[width=0.7\linewidth]{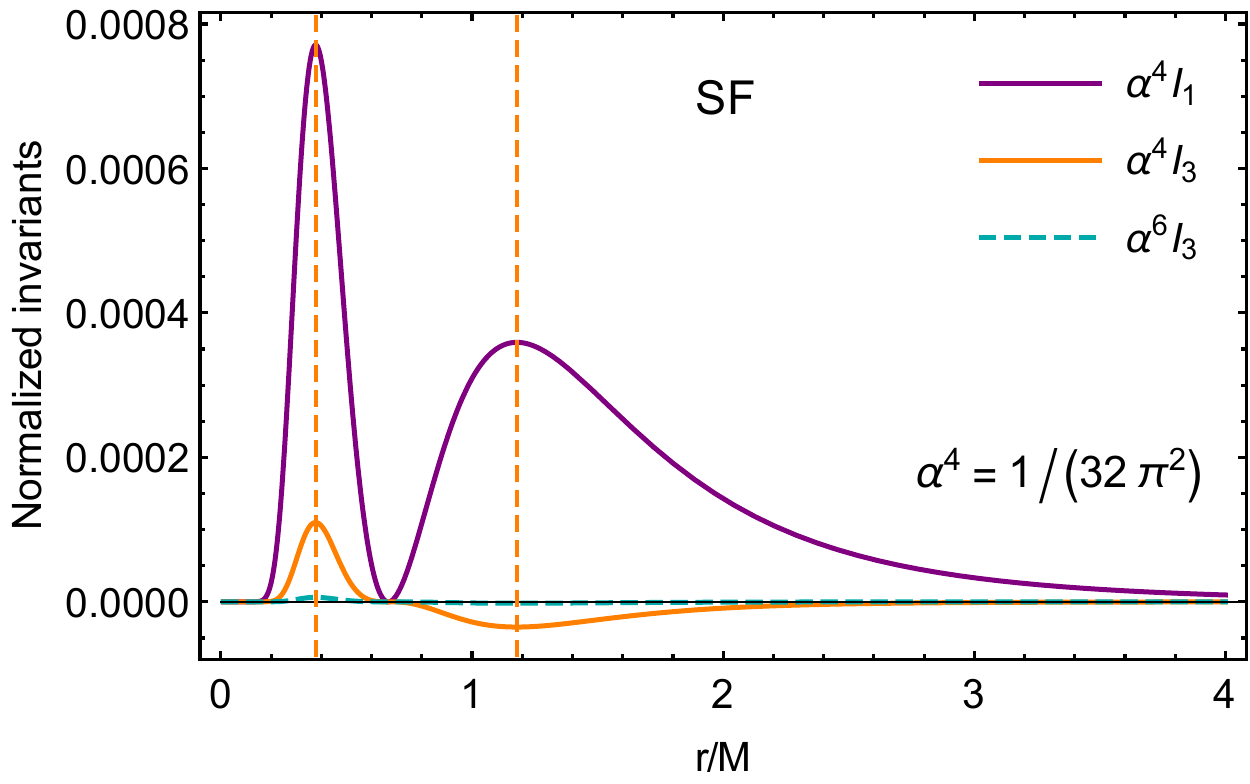}
\caption{The invariant $I_3$ has extrema (orange dashed lines) at the same coordinate distance $r$ as the Weyl invariant $W$ ($I_1$), i.e. for $r_1=0.377M$, $r_2=1.178M$ we have $I_3(r_1)=0.0347$, $I_3(r_2)=-0.0110$ (without scaling) and, in addition, $I_3$ changes sign at $r_{3}=2M/3$. The influence of this invariant (visible in $\alpha^4$ calibration, orange) in practice could be neglected, as shown in the standard $\alpha^6$ calibration (cyan, dashed).}
\label{fig:I3_fig}
\end{figure}
Now, let us consider the Weyl invariant in the exponential metric,
\begin{equation}\label{weyl}
    W \equiv I_1=C_{\alpha\beta\mu\nu}C^{\alpha\beta\mu\nu} = 
    \frac{16e^{-\frac{4M}{r}}M^2(2M-3r)^2}{3r^8},
\end{equation}
which has not one, but three extrema: two maxima at $r_{1,2}=M(7 \mp \sqrt{13})/9$, that is, $r_{1}=0.377M$, $W(r_1 ) = 0.243$ and $r_{2}=1.178M$, $W(r_2) = 0.113$, and in the interval between them there is a minimum $W(r_3)=0$ at $r_{3}=2M/3= 0.666M$.

This behavior is explained by the fact that the Weyl invariant \eqref{weyl} has a structure similar to $\mathcal{G}$ \eqref{GB}, namely, $W=K-\Xi_W$, where now
\begin{equation} \label{Xi_W}
    \Xi_W = 2R_{\alpha \beta}R^{\alpha \beta} - \frac{R^2}{3} = \frac{5R^2}{3} = \frac{20M^4 e^{-\frac{4M}{r}}}{3r^8}.
\end{equation}

However, the invariant $W$, in contrast to the topological invariant $\mathcal{G}$, does not change sign anywhere, since the components $K(r)$ and $\Xi_W(r)$ are everywhere positive and disappear asymptotically not intersecting (see Fig. \ref{fig:Weyl_fig}), but only touch each other at one point $r_3$, where the zero minimum arises, or, in the terminology of \cite{2022PDU....3500946T}, ``stationary point'' for $W$.

For completeness, retain on the remaining conformal invariants, namely \cite{Kraniotis_2022}:
\begin{equation}
    I_2=-C_{\alpha\beta}^{~~~\mu\nu}C_{\mu\nu}^{*~~\alpha\beta}=0, 
\end{equation}  
\begin{equation}   
    I_3=C_{\alpha\beta}^{~\mu\nu}C_{\mu\nu}^{~\sigma\rho}C_{\sigma\rho}^{~\alpha\beta}=\frac{32 M^3 e^{-\frac{6 M}{r}} (2 M-3 r)^3}{9 r^{12}}, 
 \end{equation}
\begin{equation}    
    I_4=-C_{\alpha\beta}^{~\mu\nu}C_{\mu\nu}^{*~~\sigma\rho}C_{\sigma\rho}^{~\alpha\beta}=0.
\end{equation} 
Two of them, containing the dual Weyl tensor, disappear in the Papapetrou metric, $I_2=I_4=0$, and the invariant $I_3$ at $r_{3}=2M/3$ changes sign - see Fig. \ref{fig:I3_fig}, which means a partial instability of the ``stationary point''.

It is obvious that there are no features in the invariants $I_1$ and $I_3$ directly associated with the critical scale $r=M$ for TWH. This conclusion is also valid for mixed invariants (see \cite{Kraniotis_2022}) that depend on both the Weyl and Ricci tensors.

\section*{Acknowledgements}
The research is funded by the Science Committee of the Ministry of Science and Higher Education
of the Republic of Kazakhstan (Programs No. BR24992807 and BR24992759).
\bibliography{0refs}
\end{document}